\documentclass[10pt]{article}
\usepackage{amsmath}
\usepackage{graphicx}
\usepackage{amsfonts}
\usepackage{amssymb}
\usepackage{epsf}
\usepackage{latexsym}

\def\80{\hspace{0.8in}}

\newcommand{\be}{\begin{enumerate}}
\newcommand{\ee}{\end{enumerate}}
\newcommand{\bi}{\begin{itemize}}
\newcommand{\ei}{\end{itemize}}
\newcommand{\bd}{\begin{description}}
\newcommand{\ed}{\end{description}}
\def\beq{\begin{equation}}
\def\eeq{\end{equation}}
\def\bea{\begin{eqnarray}}
\def\eea{\end{eqnarray}}
\def\hat{\widehat}

\def\pa{\partial}
\def\d{\textrm{d}}

\def\ttE{\mbox{\tt E}}

\def\ttL{\mbox{\tt L}}
\def\ttD{\mbox{\tt D}}

\def\ttQ{\mbox{\tt Q}}

\def\cr{\mbox{\scriptsize{\bf $\mbox{ } \times \mbox{ }$}}}

\def\md{\mbox{d}}

\def\ml{\mbox{l}}

\def\mm{\mbox{m}}
\def\mn{\mbox{n}}

\def\mD{\mbox{D}}

\def\mL{\mbox{L}}

\def\mN{\mbox{N}} 

\def\mP{\mbox{P}}

\def\sa{\mbox{\scriptsize a}}

\def\sc{\mbox{\scriptsize c}}
\def\sd{\mbox{\scriptsize d}}
\def\se{\mbox{\scriptsize e}}
\def\sf{\mbox{\scriptsize f}}
\def\sg{\mbox{\scriptsize g}}

\def\sk{\mbox{\scriptsize k}}
\def\sll{\mbox{\scriptsize l}}  
\def\sm{\mbox{\scriptsize m}}
\def\sn{\mbox{\scriptsize n}} 
\def\so{\mbox{\scriptsize o}}

\def\sr{\mbox{\scriptsize r}}
\def\sss{\mbox{\scriptsize s}}
\def\st{\mbox{\scriptsize t}}
\def\su{\mbox{\scriptsize u}}

\def\sD{\mbox{\scriptsize D}}
\def\sE{\mbox{\scriptsize E}}

\def\sG{\mbox{\scriptsize G}}

\def\sN{\mbox{\scriptsize N}}

\def\sR{\mbox{\scriptsize R}}

\def\sT{\mbox{\scriptsize T}}

\def\sY{\mbox{\scriptsize Y}}

\def\eph(B){\mbox{\scriptsize emergent(LMB)}}

\def\uRR{{\underline{R}}}
\def\uPP{{\underline{P}}}
\def\ubb{{\underline{b}}}
\def\td{\mbox{\tiny d}}

\def\tD{\mbox{\tiny D}}

\def\tN{\mbox{\tiny N}}

\def\fE{\mbox{\sffamily E}}

\def\fH{\mbox{\sffamily H}}
\def\fI{\mbox{\sffamily I}}

\def\fQ{\mbox{\sffamily Q}}
\def\fR{\mbox{\sffamily R}}
\def\fS{\mbox{\sffamily S}}
\def\fT{\mbox{\sffamily T}}

\def\fV{\mbox{\sffamily V}}
\def\fW{\mbox{\sffamily W}}

\def\sfA{\mbox{\sffamily{\scriptsize A}}}
\def\sfB{\mbox{\sffamily{\scriptsize B}}}

\def\suma{\sum\mbox{}_{\mbox{}_{\mbox{\scriptsize $i$}}}}
\def\sumi3{\sum\mbox{}_{\mbox{}_{\mbox{\scriptsize $i$=1}}}^3}
\def\sumj3{\sum\mbox{}_{\mbox{}_{\mbox{\scriptsize $j$=1}}}^3}
\def\sumk3{\sum\mbox{}_{\mbox{}_{\mbox{\scriptsize $k$=1}}}^3}

\begin{document}
\begin{titlepage}
\vspace{.7in}
\begin{center}
 
\vspace{2in} 

\LARGE{\bf QUANTUM COSMOLOGICAL METROLAND MODEL}

\vspace{0.1in}

\vspace{.4in}

\large{Edward Anderson$^{1}$ and Anne Franzen$^{2}$} 

\vspace{.2in}

\large{\em $^1$ DAMTP Cambridge U.K.}

\vspace{.2in}

\large{\em $^2$ Spinoza Institute, Utrecht, The Netherlands.}

\end{center}

\begin{abstract}

Relational particle mechanics is useful for modelling whole-universe issues such as quantum cosmology or 
the problem of time in quantum gravity, including some aspects outside the reach of comparably complex 
minisuperspace models.    
In this article, we consider the mechanics of pure shape and not scale of 4 particles on a line, so that 
the only physically significant quantities are ratios of relative separations between the constituents'  
physical objects.
Many of our ideas and workings extend to the N-particle case.
As such models' configurations resemble depictions of metro lines in public transport maps, we term 
them `N-stop metrolands'.
This 4-stop model's configuration space is a 2-sphere, from which our metroland 
mechanics interpretation is via the `cubic' tessellation.  
This model yields conserved quantities which are mathematically SO(3) objects like angular momenta 
but are physically relative dilational momenta (i.e. coordinates dotted with momenta).      
We provide and interpret various exact and approximate classical and quantum solutions for 4-stop 
metroland; from these results one can construct expectations and spreads of shape operators that admit 
interpretations as relative sizes and the `homogeneity of the model universe's contents', and also 
objects of significance for the problem of time in quantum gravity (e.g. in the na\"{\i}ve 
Schr\"{o}dinger and records theory timeless approaches).  

\end{abstract}

\vspace{3in}


\mbox{ }

\mbox{ }

\noindent$^1$ ea212@cam.ac.uk \mbox{ } , \mbox{ } $^2$ a.t.franzen@uu.nl   

\end{titlepage}

\section{Introduction}

Euclidean relational particle mechanics (ERPM) (proposed in \cite{BB82} and further studied in 
\cite{ERPM, B94I, EOT, Paris, 06I, TriCl, 08I, MGM, Cones, scaleQM, Ultra} is a mechanics in which only 
relative times, relative angles and relative separations are meaningful.  
On the other hand, in similarity relational particle mechanics (SRPM) (proposed in \cite{B03} and 
further studied in \cite{SRPM, Paris, 06II, TriCl, FORD, 08I, 08II, +Tri, Ultra}), only relative times, 
relative angles and ratios of relative separations are meaningful.  
More precisely, these theories implement the following two Barbour-type relational\footnote{RPM's are 
relational in Barbour's sense of the word rather than Rovelli's distinct one; see e.g. \cite{Rovellibook, 
B94I, EOT} for these authors' original material and \cite{08I} for a discussion of some differences.} 
postulates. 

\noindent 1) They are {\it temporally relational} \cite{BB82, RWR, Lan, Phan, FORD}, i.e. there is no 
meaningful primary notion of time for the whole system thereby described (e.g. the universe), which is  
implemented by using actions that are manifestly reparametrization invariant while also being free of 
extraneous time-related variables [such as Newtonian time or General Relativity (GR)'s lapse].   
This reparametrization invariance then directly produces primary constraints quadratic in the momenta. 

\noindent 2) They are {\it configurationally relational}, which can be conceived in terms of a certain 
group $G$ of transformations that act on the theory's configuration space $\fQ$ being held to be 
physically meaningless \cite{BB82, RWR, Lan, Phan, FORD, Ultra}.   
This can be implemented by such as 
using arbitrary-$G$-frame-corrected quantities rather than `bare' $\fQ$-configurations.
For, despite this augmenting $\fQ$ to the principal bundle $P(\fQ, G)$, variation with respect to each 
adjoined independent auxiliary $G$-variable produces a secondary constraint linear in the momenta 
which removes one $G$ degree of freedom and one redundant degree of freedom among the $\fQ$ variables. 
Thus, one ends up dealing with the desired reduced configuration space -- the quotient space $\fQ/G$.  
Configurational relationalism includes as subcases both spatial relationalism (for spatial transformations) 
and internal relationalism (in the sense of gauge theory).

For ERPM, the Jacobi-type \cite{Lanczos} action is\footnote{$\uRR^i$ 
are relative Jacobi coordinates \cite{Marchal}: linear combinations of relative particle separation 
vectors that produce a diagonal kinetic term and are particular inter-particle cluster separation 
vectors with associated cluster masses $\mu_i$. 
Lower-case Latin indices run over 1 to n = N -- 1 for N the number of particles, and lower-case greek 
ones are spatial indices; the spatial dimension is $d$.   
$\lambda$ is label time and $\dot{\mbox{ }}$ is the derivative with respect to this.   
Using such relative coordinates, one has already incorporated the highly trivial translation part of the 
Euclidean or similarity groups.
$\dot{\ubb}$ is a rotational auxiliary velocity whereby the rotation part of these groups is implemented. 
In the SRPM case, $\dot{c}$ is a dilational auxiliary velocity implementing the additional scaling part.
$M_{i\alpha j\beta} = \mu_i\delta_{ij}\delta_{\alpha\beta}$ is the mass matrix with determinant $M$ and 
inverse $N^{i\alpha j\beta}$.
$\uPP_i$ is the momentum conjugate to $\uRR^i$.
$I$ is the moment of inertia, $\sum_i\mu_i|\uRR^i|^2$.  
For ERPM, $T$, $V$ and $E$ are kinetic, potential and total energy terms with the usual physical 
dimensions.  
In our `pure shape' formulation of the SRPM, the kinetic term $\fT$ has dimensions of 
(energy)/$I$ and $\fE - \fV$ has dimensions of (energy)$\times I$.  
Consistency dictates that this $\fV$ additionally be a homogeneous function of the $\uRR^i$; in fact, 
in the given `pure shape' formulation, it must be homogeneous of degree zero. 
This turns out not to be a heavy restriction due to $I$ being constant after variation and useable to 
homogenize (see Sec 2 for an example).
While an actual energy is prohibited by this consistency, the abovementioned constant $\fE$ is 
permissible instead.}
\beq
\fI = 2\int\d\lambda\sqrt{T\{ E - V\}} 
\mbox{ } , \mbox{ } \mbox{ with }  
T = \suma\mbox{ } \mu_i \{\dot{\uRR}\mbox{}^i - \dot{\ubb} \cr \uRR^i\}^2/2  
\label{1} \mbox{ } ,
\eeq
and for SRPM, our presentation of it is 
\beq
\fI = 2\int\d\lambda\sqrt{\fT\{\fE - \fV\}} 
\mbox{ } , \mbox{ } \mbox{ with }
\fT = \suma\mbox{ } \mu_i\{\dot{\uRR}\mbox{}^i - \dot{\ubb} \cr \uRR^i + \dot{c}\uRR^i\}^2/2I   
\label{2} \mbox{ } .  
\eeq 
These implement the above relational postulates for the corresponding Euclidean and similarity $G$'s 
\cite{BB82, B03, 06I, 06II, 08I}.
Equivalent theories formulated directly in terms of rotational (and dilational) invariant quantities 
can also be arrived at by considering the space of shapes and mechanics thereupon \cite{Kendall, FORD}.  
It is then of interest what structure one gets when one quantizes such theories \cite{SRBS, 06I, SemiclI, 
08II, MGM, Gryb, Cones, +Tri, scaleQM, Ultra, 08III, SemiclIII}.      


The Barbour-type indirect formulation of RPM's (\ref{1},\ref{2}), moreover, is particularly interesting 
through how the geometrodynamical form of GR can be cast in direct parallel: it also obeys postulates 
1) and 2) implemented as follows \cite{RWR, Phan, FEPI} (some features of which are already anticipated 
in \cite{BSW}).\footnote{The spatial topology $\Sigma$
is taken to be compact without boundary. 
$h_{\mu\nu}$ is a spatial 3-metric thereupon, with determinant $h$, covariant derivative $D_{\mu}$, 
Ricci scalar Ric($h$) and conjugate momentum $\pi^{\mu\nu}$.  
$\Lambda$ is the cosmological constant.
${\cal M}^{\mu\nu\rho\sigma} = h^{\mu\rho}h^{\nu\sigma} - h^{\mu\nu}h^{\rho\sigma}$ is the inverse 
DeWitt supermetric with determinant ${\cal M}$ and inverse ${\cal N}_{\mu\nu\rho\sigma}$. 
To represent this as a configuration space metric (i.e. with just two indices, and 
downstairs), use DeWitt's 2 index to 1 index map \cite{DeWitt}.
$\dot{F}^{\mu}$ is the velocity of the frame; in the manifestly relational formulation of GR, this 
cyclic velocity plays the role more usually played by the shift Lagrange multiplier coordinate.
$\pounds_{\dot{F}}$ is the Lie derivative with respect to $\dot{F}^{\mu}$.} 
\beq
\fS^{\sG\sR} = 
2\int\d\lambda\int\d^{3}x\sqrt{h}\sqrt{\fT^{\sG\sR}_{}
\{\mbox{Ric}(h) - 2\Lambda\}} \mbox{ } \mbox{  for } \mbox{ }   
\fT^{\sG\sR}_{} = 
\frac{1}{4}{\cal M}^{\mu\nu\rho\sigma}\{\dot{h}_{\mu\nu} - \pounds_{\dot{F}}h_{\mu\nu}\}
                                      \{\dot{h}_{\rho\sigma} - \pounds_{\dot{F}}h_{\rho\sigma}\}
\label{GRaction} \mbox{ } ; 
\eeq
in this case, $\fQ$ is the space Riem($\Sigma$) of Riemannian 3-metrics on a fixed spatial topology 
$\Sigma$, and $G$ is the corresponding 3-diffeomorphism group, Diff($\Sigma$).  

\mbox{ }

The way that the physical equations follow from each of the above actions then has many parallels.
By reparametrization invariance \cite{Dirac} each has a primary constraint quadratic in the momenta: 
for GR the Hamiltonian constraint
\beq
{\cal H} \equiv {\cal N}_{\mu\nu\rho\sigma}\pi^{\mu\nu}\pi^{\rho\sigma} - \sqrt{h}\{\mbox{Ric}(h) - 
2\Lambda\} = 0 
\eeq
and, for ERPM and SRPM respectively, the `energy constraints':
\beq
\ttQ \equiv N^{i\alpha j\beta}P_{i\alpha}P_{j\beta}/2 +  V =  E \mbox{ } \mbox{ } , \mbox{ } \mbox{ } 
\ttQ \equiv I N^{i\alpha j\beta}P_{i\alpha}P_{j\beta}/2 + \fV = \fE \mbox{ }. \mbox{ }
\eeq
By variation with respect to the auxiliary $G$-variables, each relational theory has constraints linear in the 
momenta: for GR, the momentum constraint 

\noindent
\beq
{\cal L}_{\mu} \equiv - 2D_{\nu}{\pi^{\nu}}_{\mu} = 0 \mbox{ } 
\label{GRmom}
\eeq
from variation with respect to $F^{\mu}$, and, for RPM's, the zero total angular momentum and zero total dilational 
momentum constraints 
\beq
\underline{\ttL} \equiv \suma\mbox{ } \uRR^i \cr \uPP_i = 0 
\mbox{ } \mbox{ } , \mbox{ } \mbox{ } 
\ttD \equiv \suma\mbox{ } \uRR^i \cdot \uPP_i = 0 
\label{LR}
\eeq
from variation with respect to $b^{\mu}$ and $c$ (so the latter constraint occurs only in SRPM).  
The zero total dilational momentum constraint moreover closely parallels the well-known GR maximal 
slicing condition \cite{Lich}, $h_{\mu\nu}\pi^{\mu\nu} = 0$.  
Furthermore, much like generalizing maximal slicing to constant mean curvature slicing \cite{York72} 
turns on a `York time' variable \cite{K81, K92} $t^{\sY\so\sr\sk} \equiv {2}h_{\mu\nu}
\pi^{\mu\nu}/3\sqrt{h} = c(\lambda \mbox{ alone})$, one can think of the passage from SRPM to ERPM as 
involving an extra `Euler time' variable $t^{\sE\su\sll\se\sr} \propto \suma \uRR^i \cdot \uPP_i = 
C(\lambda)$. 
This is all underlied for both GR and RPM's by shape-scale splits, the role of scale being 
played by $\sqrt{I}$ or $I$ for RPM's and by such as the scalefactor $a$ or $\sqrt{h}$ in GR.  
In both cases it is then tempting to use the singled-out scale as a time variable but this runs into 
monotonicity problems which are avoided by using as times the quantities conjugate to (a function of) 
the scale, i.e. $t^{\sY\so\sr\sk}$ and $t^{\sE\su\sll\se\sr}$.

There are further analogies at the configuration space level.    
If 1) $\fR(\mN, d)$ the {\it relative space} of relative interparticle (cluster) separation vectors and 
Riem($\Sigma$) are held to be analogous, then so are    
2) {\it Relational space} = $\fR(\mN, d)$/Rot($d$) 
for Rot($d$) the $d$-dimensional rotations and 
superspace($\Sigma$) = Riem($\Sigma$)/Diff($\Sigma$).  
3) {\it Shape space} = $\fR(\mN, d)$/Rot $\times$ Dil for Dil the dilations and 
conformal superspace \cite{York74, ABFO} 
CS($\Sigma$) = Riem($\Sigma$)/Diff($\Sigma$) $\times$ Conf($\Sigma$) 
for Conf($\Sigma$) the conformal transformations on $\Sigma$.
4) The cone representation of relational space in shape-scale split variables \cite{Cones} and 

\noindent 
$\{$CS + V$\}$($\Sigma$) = Riem($\Sigma$)/Diff($\Sigma$) $\times$ VPConf($\Sigma$) \cite{York72} for 
VPConf($\Sigma$) the conformal transformations that preserve the volume of the universe, V  
\cite{ABFKO}.  
Also, both these GR and RPM configuration spaces are in general stratified, and both have physically 
significant bad points (e.g. $a = 0$ is the Big Bang and $I = 0$ is the maximal collision).

\mbox{ }

There are yet more analogies \cite{Battelle, DeWitt, K81, PW83, K91, K92, B94I, B94II, K99, EOT, 
Kieferbook, 06II, Smolin08}  at the level of various strategies toward the resolution of the Problem of 
Time\footnote{This notorious 
problem occurs because `time' takes a different meaning in each of GR and ordinary quantum theory.  
This incompatibility underscores a number of problems with trying to replace these two branches with a 
single framework in situations in which the premises of both apply, namely in black holes and in the 
very early universe.  
One facet of the Problem of Time that shows up in attempting canonical quantization is that the lack of 
linear momentum dependence of the GR Hamiltonian constraint, but it does have many other facets 
\cite{K92}.}
and various other aspects of quantum cosmology.
The above quadratic constraints give frozen (i.e. timeless or stationary) quantum equations.  
For GR, this is the Wheeler--DeWitt equation,
\beq
\hat{\cal H}\Psi =
- {\hbar^2}
`\left\{
\frac{1}{\sqrt{{\cal M}}}\frac{\delta}{\delta h^{{\mu\nu}}}
\left\{
\sqrt{{\cal M}}{\cal N}^{\mu\nu\rho\sigma}\frac{\delta\Psi}{\delta h^{{\rho\sigma}}}
\right\} - \xi \,\mbox{Ric}({\cal M})\Psi\right\}\mbox{'}
-  \sqrt{h}(\mbox{Ric}(h) - 2\Lambda   \}\Psi  
= 0
\label{WDE} \mbox{ } ,   
\eeq
where $\Psi$ is the wavefunction of the universe; `\mbox{ }' implies in general various well-definedness 
issues (see e.g. \cite{08II} for a summary) and need for a choice of operator-ordering (we use conformal 
ordering in this paper, c.f. Sec 3.1).   
Correspondingly, for RPM's, 
\beq
\hat{\ttQ}\Psi = 
-\frac{\hbar^2}{2}
\left\{ 
\frac{1}{\sqrt{M}}\frac{\pa}{\pa Q^{\sfA}}
\left\{
N^{\sfA\sfB}\sqrt{M}\frac{\pa}{\pa Q^{\sfB}}
\right\} - \xi \mbox{Ric}(M)
\right\}\Psi + \fV\Psi = \fE\Psi \mbox{ } .  
\label{RPMWDE}
\eeq  
For the moment $\sfA = i\alpha$, $\sfB = j\beta$ and the $Q^{\sfA}$ are the $R^{i\alpha}$.  
$N^{\sfA\sfB}$ is the inverse mass matrix for ERPM and $I$ times it for SRPM.  
However, we use this equation more generally than that below for reduced RPM's in which linear 
constraints have been taken care of, this being explicitly possible in 1- or 2-$d$ \cite{FORD}.

An important feature of GR (and one missed out by minisuperspace models \cite{Magic, Mini}) is that of 
linear constraints causing substantial complications e.g. in attempted resolutions of the Problem of 
Time.  
That is the momentum constraint for GR [(\ref{GRmom}) or its quantum counterpart], while RPM's have the 
linear constraints [(\ref{LR}) or their quantum counterparts].  
However, minisuperspaces, unlike RPM's, have more specific and GR-inherited potentials and indefinite 
kinetic terms.  
Thus both minisuperspace and RPM's are valuable in complementary ways as toy models.\footnote{Midisuperspace 
\cite{Midi} unites all these desirable features but is unfortunately then calculationally too hard for 
many of the strategies.}  

Some of the strategies toward resolving the Problem of Time are as follows.  

\noindent A) 
Perhaps one is to find a time hidden within classical GR \cite{K92} and thus obtain a wave equation that depends on 
it from the outset at the quantum level.   
York time is a GR example of such and Euler time is an ERPM model of it.    

\noindent B)
Perhaps one has slow, heavy `H'  variables that provide an approximate timestandard with respect to which 
the other fast, light `L' degrees of freedom evolve \cite{HallHaw, K92, Kieferbook}.  
In quantum cosmology the role of H is played by scale (and homogeneous matter modes), so ERPM's in 
scale--shape split are more faithful semiclassical models of this than SRPM's themselves can muster.  

\noindent C) A number of approaches take timelessness at face value. 
One considers only questions about the universe `being', rather than `becoming', a certain way.  
This can cause at least some practical limitations, but nevertheless can address at least {\sl some} 
questions of interest. 
For example, Hawking and Page's {\it na\"{\i}ve Schr\"{o}dinger interpretation} \cite{HP86UW89} 
concerns the `being' probabilities for universe properties such as: what is the probability that the 
universe is large? 
Flat? 
Isotropic? 
Homogeneous?   
One obtains these via consideration of integrals of $|\Psi|^2$ over suitable regions of the 
configuration space. 
This approach is termed `na\"{\i}ve' due to it not using any further features of the constraint 
equations.  
The {\it conditional probabilities interpretation} \cite{PW83} goes further by addressing conditioned 
questions of `being' such as `what is the probability that the universe is flat given that it is 
isotropic'?  
{\it Records theory} \cite{PW83, GMH, B94II, EOT, H99, Records} involves localized subconfigurations 
of a single instant -- whether these contain useable information, are correlated to each other, 
and whether a semblance of dynamics or history arises from this.  
RPM's are superior to minisuperspace for such a study as they have a notion of localization in space, 
and more options for well-characterized localization in configuration space through their kinetic 
terms possessing positive-definite metrics.  

\noindent D) Perhaps instead it is the histories that are primary ({\it histories theory} \cite{GMH, 
Hartle}).    
There is a records theory within histories theory, and histories decohereing is one possible way of 
obtaining a semiclassical regime in the first place, making B) to D) of particular interest to one of us 
\cite{Records, MGM}.     

\noindent E) Distinct timeless approaches involve {\it evolving constants of the motion} (`Heisenberg' 
rather than `Schr\"{o}dinger' style QM), or {\it partial observables} \cite{Rovellibook} (which are used 
in Loop Quantum Gravity's {\it master constraint program} \cite{Thiemann}).  

\noindent
Some approaches to the Problem of Time that do {\sl not} have an RPM analogue include superspace time 
(which requires indefinite configuration spaces) and third quantization (which requires field theoretic 
rather than finite models).

\mbox{ }  

\noindent We are in the process of building up a reasonable set of RPM models, paralleling e.g. the 
development of minisuperspace in the early 70's \cite{Magic, Mini}, or Carlip's work in the 90's for 2 + 
1 gravity \cite{Carlip} (see \cite{K92} for yet further useful toy model arenas for Problem of Time  
approaches). 
Also RPM's serve as a bridge from highly-studied ideas in molecular physics and `mini- and 
midi'superspace, which may serve to import ideas and tools from the former to the latter.

Our build-up is for RPM's in 1-$d$ and 2-$d$; for N particles, we term these, respectively, 
{\it N-stop metroland} and {\it N-a-gonland} (the first two nontrivial N-a-gonlands we furthermore refer 
to as {\it triangleland} and {\it quadrilateralland}). 
We choose to study these models because their configuration spaces are highly tractable mathematically 
\cite{Kendall, FORD}: $\mathbb{S}^{\sN - 2}$ spheres in 1-$d$ and $\mathbb{CP}^{\sN - 2}$ complex 
projective spaces in 2-$d$.
This is for the choice of plain shapes rather than oriented shapes. 
[I.e. we make the choice of treating 
each shape and its mirror image as distinct; in this paper's 4-stop metroland model, this means that we 
regard the 1,2,3,4 ordering of the particles to be distinct from the 4,3,2,1 one. 
The opposite choice gives configuration spaces $\mathbb{S}^k/\mathbb{Z}_2 = \mathbb{RP}^k$ (real 
projective spaces), and $\mathbb{CP}^k/\mathbb{Z}_2$ 
%
%
which are somewhat harder to model.]  
N.B. that the interesting theoretical parallels between GR and RPM's are unaffected by our choice of 
plain shapes and of low-$d$ RPM's.   
We are presently studying scalefree models as these are more straightforward than models with scale 
(though we will need to move on to scaled models as regards reasonably quantum-cosmologically realistic 
modelling of the semiclassical approach; note also that scalefree problems occur as a subproblems in 
models with scale \cite{BGS03, FORD, 08I}, so studying these first also makes sense even from this 
semiclassical quantum cosmological perspective).

This paper considers scalefree 4-stop metroland (the smallest scalefree metroland to have the 
nontrivialities associated with having 2 physical degrees of freedom, so that one physical quantity can 
be expressed in terms of another, a feature necessary for records theory's correlations, while  
semiclassical approaches need at least one H and at least one L, decoherence only makes sense if one 
thing decoheres another, and so on).
Additionally, there are indications that 4-stop metroland is simpler than triangleland \cite{Cones} 
(which also has two physical degrees of freedom and spherical shape space by $\mathbb{CP}^1 = 
\mathbb{S}^2$), particularly in the cases with scale and at the quantum level, so that the present paper 
is useful toward how to subsequently deal with these other more complicated cases.  
Also, many of the present paper's workings readily extend to N-stop metroland.  
4-stop metroland and triangleland are both useful preliminaries for studying quadrilateralland, which is 
the simplest RPM to exhibit a number of geometrical nontrivialities, including some relevant to Problem 
of Time approaches and some that are archetypal of 2-$d$ problems in ways that triangleland is not.  
Moreover, 4-stop metroland itself is already suitable for study of various timeless approaches to 
the Problem of Time.  

\mbox{ }

In Sec 2 we begin with a classical treatment of 4-stop metroland in its reduced form.  
We give a tessellation of the shape sphere corresponding to 4-stop metroland's physical interpretation 
and provide useful and meaningful shape quantities for our study.   
We then study 4-stop metroland's equations of motion and its conserved quantities, among which some have 
angular momentum-like mathematics but are physically dilational rather than angular momenta.  
We interpret simple subcases of multiple harmonic oscillator-like potentials' solutions using our tessellation 
and shape quantities, concentrating on the case of two localized and well-separated subsystems that is 
motivated by our interest in timeless approaches.

In Sec 3 we consider time-independent Schr\"{o}dinger equations for these problems, 
by firstly interpreting their exact and asymptotic solutions against our tessellation. 
Secondly, we compute expectations and spreads of our shape quantities promoted to quantum-mechanical 
shape operators.   
Thirdly, we consider perturbations about the simplest case in which the `springs' balance each other 
out to produce a constant potential and hence spherical harmonics mathematics.    
Fourthly, we note and apply a number of analogies with molecular physics to our study.  
We conclude in Sec 4, including discussion of how our model and slightly larger versions thereof can be 
used as an arena for investigation of a number of Problem of Time strategies -- of which we provide 
na\"{\i}ve Schr\"{o}dinger interpretation examples -- and by commenting on `mini- and 
midi'superspace counterparts of this paper's shape operators.

\section{4-stop metroland at the classical level}

\subsection{Passage to reduced form of 4-stop metroland and useful coordinatizations of it}

The unreduced action is given by the SRPM case within eq. (2) further restricted to being in 
1-$d$ (so there are no rotations) and for 4 particles, and so 3 relative separations and thus 3 relative 
Jacobi coordinates, $R^i$: 
\beq
\fI = 2\int\d\lambda\sqrt{{\fT}\{\fE - \fV\}} \mbox{ } \mbox{ with } \mbox{ } 
\fT = \sumi3\mu_i\{\dot{R}^i + \dot{c}R^i\}^2/2I \mbox{ } .  
\eeq
We will find it more convenient to deal with the subsequent physics in terms of $\iota^{i} \equiv 
\sqrt{\mu_i}R^i$ the {\sl mass-weighted} relative Jacobi coordinates [which are physically the square 
roots of the partial moments of inertia  $I^i = \mu_iR^{i\,2}$ (no sum)], and, after variation, 
their `{\sl normalized}' counterparts $n^i = \iota^i/\iota$ and $N^i = I^i/I$ for $I$ the total moment 
of inertia and $\iota = \sqrt{I}$.  
It will often be convenient to use $n_x$, $n_y$, $n_z$ for the components of $n^i$.  
We take these Jacobi coordinates to be, in terms of particle position coordinates, Jacobi H-coordinates 
rather than Jacobi K-coordinates (Fig 1) with quantum cosmological and records theoretic applications 
in mind: two equal particle number clusters treated on the same footing, each could model the seed of a 
galaxy, or be a nontrivial record (of which we need at least 2 to consider correlations between records).  

{            \begin{figure}[ht]
\centering
\includegraphics[width=0.8\textwidth]{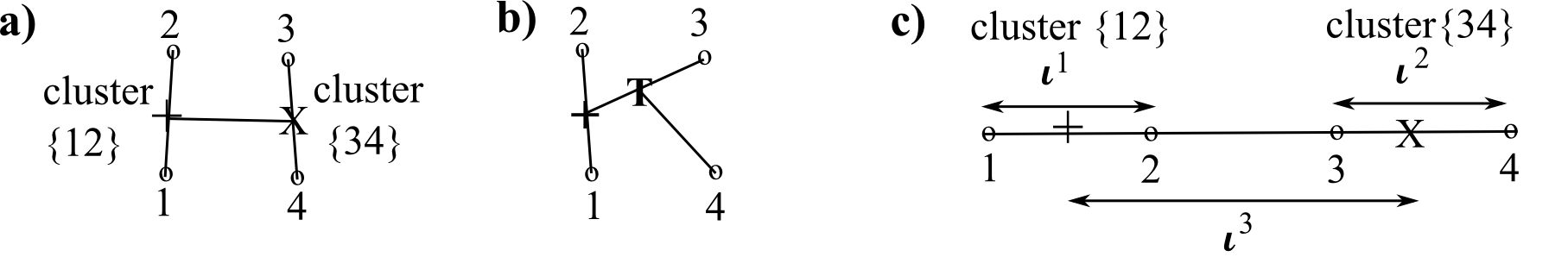}
\caption[Text der im Bilderverzeichnis auftaucht]{        \footnotesize{    a) and b) explain in 2-$d$ 
the origin of the names H- and K-coordinates.   
Using $\{a...c\}$ to denote the cluster composed of particles a, ..., c ordered from left to right, + is 
the centre of mass (COM) of cluster \{12\}, X is the COM of cluster \{34\} and T is the COM of the 
triple cluster \{123\}.  
c) What H-coordinates look like in 1-$d$: the H has been `squashed'.    }        }
%
\label{Fig1}\end{figure}           }

Now let us perform some variational manoeuvres on the above action. 
It is useful to bear in mind from the outset that our 
%
%
4-stop metroland's reduced configuration space is 
$\mathbb{S}^2$ and we are trying to bring this out as cleanly as possible by removing extraneous variables and 
seeking for standard coordinates on this.
Variation with respect to the dilational auxiliary $c$ gives the dilational constraint (\ref{LR}), 
the Lagrangian form for which can be rearranged to 
\beq
\dot{c} = 
-\sumi3 \mu_i R^i\dot{R}^i 
\left/ 
\sumj3 \mu_j\{R^j\}^2 \right. \mbox{ } ,
\eeq
and used to eliminate $\dot{c}$  from the action, producing (2) but with $\fT_{\sr\se\sd}$ in 
place of $\fT$:
\beq
\fT_{\sr\se\sd} = 
\left.
\big\{
\sumi3\{\iota^i\}^2\sumj3\{\dot{\iota}^j\}^2 - 
\big\{
\sumi3\iota^i\dot{\iota}^i
\big\}^2
\big\}
\right/
2\big\{\sumk3 \{\iota^k\}^2\big\}^2
\label{iaction} \mbox{ } .  
\eeq
Then, via the coordinate transformation
\beq
\Theta = \mbox{arctan}\left(\sqrt{\{\iota^1\}^2 + \{\iota^2\}^2}/\iota^3\right) 
\mbox{ } , \mbox{ } 
\Phi = \mbox{arctan}\left({\iota^2}/{\iota^1}\right) \mbox{ } , 
\label{Var}
\eeq
(\ref{iaction}) becomes
\beq
\fT_{\mathbb{S}^2} = \{\dot{\Theta}^2 + \mbox{sin}^2\Theta\,\dot{\Phi}^2\}/2 \mbox{ } .
\label{Tsp}
\eeq
The coordinate ranges are $0 < \Theta < \pi$ and $0 \leq \Phi < 2\pi$, so these are geometrically the 
standard azimuthal and polar spherical angles on the unit shape space sphere $\mathbb{S}^2$.    
Inversely,
\beq 
\iota^1 = \iota \,\mbox{sin}\,\Theta\,\mbox{cos}\,\Phi \mbox{ } , \mbox{ } 
\iota^2 = \iota \,\mbox{sin}\,\Theta\,\mbox{sin}\,\Phi \mbox{ } , \mbox{ } 
\iota^3 = \iota \,\mbox{cos}\,\Theta               \mbox{ } . \mbox{ } 
\eeq

Thus 4-stop metroland has $\iota \equiv \sqrt{I}$ playing a (constant) radius role, and the $\iota^i$ 
are Cartesian coordinates in the `surrounding' Euclidean relational space $R(4, 1) = \mathbb{R}^3$, 
subject to the on-sphere condition 
\beq
\sumi3 I^i = 
\sumi3 \{\iota^i\}^2 = \{\iota\}^2 = I \mbox{ }  \mbox{ } \mbox{ (constant) } , \mbox{ } 
\mbox{ or } \mbox{ } 
\sumi3 N^i = \sumi3 \{n^i\}^2 = 1 \mbox{ } .  
\label{New19}
\eeq
The $n^i$ are then the components of the unit Cartesian vector [$(\mbox{sin}\,\Theta\,\mbox{cos}\,\Phi, 
\mbox{sin}\,\Theta\,\mbox{sin}\,\Phi, \mbox{cos}\,\Theta)$ in spherical polar coordinates].
This should be contrasted with the way the sphere arising in scalefree triangleland \cite{08I} being 
harder to deal with from the perspective of the `surrounding' Euclidean relational configuration space, 
which is $\mathbb{R}^3$.  
Scalefree triangleland's $\underline{\iota}^1$ and $\underline{\iota}^2$ are related to the Cartesian 
coordinates of the surrounding relational space not in the above familiar Cartesian way, but rather in 
the less straightforward `Dragt' way \cite{Cones}, corresponding to having to use not $\iota$ but $I$ as 
radial variable.

The following formulae are also useful below:  

\noindent
\beq
\mbox{cos}\,\Theta = {n_z} , \mbox{ } 
\mbox{sin}\,\Theta = \sqrt{{n_x}\mbox{}^2 + {n_y}\mbox{}^2} , \mbox{ }
\mbox{cos}\,\Phi = \frac{n_x}{\sqrt{{n_x}^2 + {n_y}^2}} , \mbox{ } 
\mbox{sin}\,\Phi = \frac{n_y}{\sqrt{{n_x}^2 + {n_y}^2}} , \mbox{ } 
\mbox{cos}\,2\Phi = \frac{{n_x}^2 - {n_y}^2}{{n_x}^2 + {n_y}^2} , \mbox{ }
\mbox{sin}\,2\Phi = \frac{2n_xn_y}{{n_x}^2 + {n_y}^2}  ,   
\label{useful}
\eeq 
and what is (from the geometrical perspective) a stereographic radial coordinate,    
\beq
{\cal R} = \mbox{tan$\frac{\Theta}{2}$} = \sqrt{\{1 - n_z\}/\{1 + n_z\}} \mbox{ } .
\label{Rdef}
\eeq

\subsection{Action and banal-conformal representations for this paper}

We take the Jacobi action corresponding to (\ref{Tsp}), 
\beq
\fI = 2\int\d\lambda\sqrt{\fT_{\mathbb{S}^2}\{\fE - \fV\}}
\eeq 
to be primary.  
As well as by Sec 2.1's reduction, this can be obtained by \cite{FORD} considering a natural mechanics 
in Jacobi's geometrical sense \cite{Lanczos} on the space of shapes \cite{Kendall}.

The `banal' conformal invariance of each of the above product-type actions $\fT \longrightarrow \Omega^2 
\fT$, $\fE - \fV \longrightarrow \{\fE - \fV\}/\Omega^2$ is useful below in `passing' factors between 
$\fT$ and $\fE - \fV$ (which we term `picking a distinct banal representation').  
The above-given forms of $\fT$ and $\fE - \fV$ are the geometrically-natural ones (both in the 
scale-invariant sense and in the sense of having the standard spherical metric on the shape space sphere)
and  mechanically-natural in the sense that $\fE$ itself appears in them rather than $\fE$ times some 
power of the moment of inertia.  
However, in some applications, (${\cal R}$, $\Phi$) coordinates and $\Omega = \{1 + {\cal R}^2\}/2$ is 
useful; we term this the `flat banal representation' as its $\fT$ is flat, and denote it by 
tilde-ing.


{            \begin{figure}[ht]
\centering
\includegraphics[width=1\textwidth]{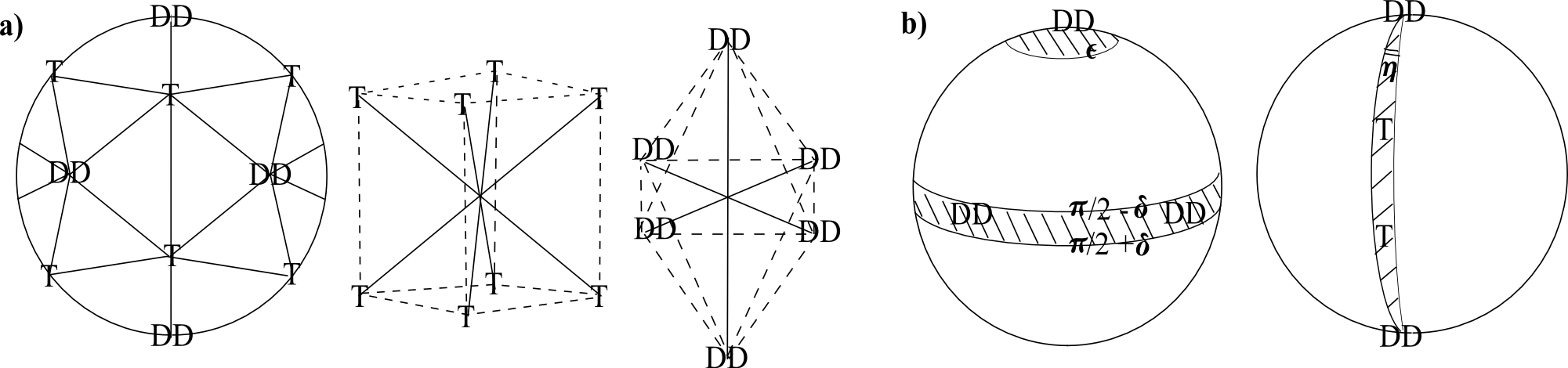}
\caption[Text der im Bilderverzeichnis auftaucht]{        \footnotesize{    a) On the configuration space 
represented as a sphere, there are 8 triple collision (T) points and 6 double-double (DD) collision points. 
Each DD is attached to 4 T's, and each T to 3 T's and 3 DD's, in each case by single double collision 
lines.  
This forms a tessellation with 24 identical spherical isosceles triangle faces, 36 edges and 14 vertices.
The T's and DD's form respectively the vertices of a cube and the octahaedron dual to it (dashed lines 
in the second and third subfigures), so that the physical interpretation has the symmetry group of the 
cube, of order 48.  
This is isomorphic to $S_4 \times \mathbb{Z}_2$ for $S_4$ the permutation group of 4 objects, thus 
realizing the ways of labelling the 4 particles and ascribing an orientation.    
See p 72-75 of \cite{Magnus} for mathematical discussion of this tessellation and \cite{Tessi} for 
another occurrence of it in mechanics.
In this arrangement, the T's and DD's form 7 antipodal pairs, thus picking out 7 preferred axes.  
The 3 axes corresponding to antipodal DD pairs are related to the 3 permutations of Jacobi H-coordinates
and 
the 4 axes corresponding to antipodal T  pairs are related to the 4 permutations of Jacobi K-coordinates.
This relation is in the sense that the poles in each case correspond to what each coordinatization picks 
out as intra-cluster coordinates both going to zero i.e. collapse of both clusters for an H or collapse 
of the triple cluster for a K.  

\noindent
b) To make statements concerning shapes being near a DD or T -- i.e. {\it well-localized} (intra-cluster 
distances far smaller than external distances to non-member particles/clusters), spherical caps $\Theta 
\leq \epsilon$ are useful in the corresponding spherical polar coordinate chart. 
In particular, with this paper's usual choice of axis, the polar caps are where there is both \{12\} and 
\{34\} localization, so following clusters \{12\} and \{34\} makes sense.    
Denote this clustering (i.e. partition into clusters) by \{12,34\}.  
The opposite notion is {\it merged clusters} for which the COM's of \{12\} and \{34\} are near each 
other so that these clusters largely overlap (which is, in a certain sense, a more `homogeneous' universe 
model).
This corresponds to belts $\pi/2 - \delta \leq \Theta \leq \pi/2 + \delta$ around the equator. 
(Multi)lunes $\Phi_0 - \eta \leq \Phi \leq \Phi_0 + \eta$ also correspond 
to physically meaningful statements. 
E.g. being in the bilune around the Greenwich meridian means that the \{34\} cluster is localized, being 
in the bilune around the `Bangladeshi' meridian perpendicular to the Greenwich one means that the \{12\} 
cluster is localized and being in the tetralune at $\pi/4$ to all of these signifies that clusters 
\{12\} and \{34\} are of similar size', i.e. $\eta$-close to {\it contents homogeneity} (i.e. that the 
particle clusters that make up the model universe are, among themselves, of similar constitution).
The above sort of approximate notions of shape are in the spirit of those used in e.g. Kendall et al. 
\cite{Kendall}, and we make use of the corresponding configuration space regions in our na\"{\i}ve 
Schr\"{o}dinger approach calculations in Sec 4.3.
Moreover, note that notions of `relative size' and `similar contents' here in fact involve more concretely 
the $\sqrt{\mbox{mass}} \times$ distance combination (whose squares are partial moments of 
inertia).    }        }
\label{Fig2}\end{figure}            }

\subsection{Physical interpretation by tessellation, charts and shape operators}

The Jacobi H-coordinates in use are better-adapted for `seeing' double-double collisions [see Fig 2a)] 
rather than triple collisions (the opposite is the case for Jacobi K-coordinates), so that it is useful 
to preliminarily work out and graphically 
represent the mechanical interpretations of the various zones 
of our problem's configuration space of shapes: Fig 2.    
This

\noindent is of considerable use below in interpreting 
classical trajectories (as paths upon this figure) and classical potentials and quantum-mechanical 
probability density functions (as height functions over this figure).    
Spherical polar coordinates about each axis in Fig 2 are natural for the study of the corresponding H 
or K structure.  
Thus each choice of H- or K-coordinates has a different natural spherical polar coordinate chart.  
Any two of these natural charts suffice to form an atlas for the sphere (each goes bad solely at its 
poles, where its axial angle ceases to be defined).  
To look extremely close to a pole, one can `cartesianize' e.g. after projecting the relevant hemisphere 
onto the equatorial disc.


The $n^i$ are interesting quantities with which to describe the shape of the configuration. 
Of course, only 2 of the 3 $n^i$ are independent, by the on-sphere restriction (\ref{New19}).  
For the example of H-coordinates that follows the \{12,34\} clustering arrangement that we follow in 
particular in this paper, the corresponding $n_z$ is a signed 
quantifier of the relative size of the universe from the perspective of an observer in either cluster.  
On the other hand, $n_x$ is a signed quantifier of the size of the universe from the less Copernican 
perspective of an observer specifically in cluster \{12\} [$n_y$ has the same meaning but for cluster 
\{34\}].    
Thus we term these shape quantities RelSize(12,34), RelSize(1,2) and RelSize(3,4) 
respectively.\footnote{There are 
other such operators corresponding to attaching significance to other clustering arrangements obtained by 
permuting the particles in defining the Jacobi H-coordinates, while similar quantities can be defined 
for the various permutations of Jacobi K-coordinates; see \cite{+Tri} for a brief account of the shape 
operators for each of these.
RelSize(12,34) is also $\sqrt{1 - {n_x}^2 - {n_y}^2}$, so it can also be viewed as a `measure' of 
noncollapse of at least one of the model universe's clusters; one can readily work out such `dual 
statements' for other shape quantities.}  

RelSize(12,34) small means that clusters \{12\} and \{34\} are merged, and corresponds 
geometrically to the equatorial belt.  
RelSize(12,34) large means physically that clusters \{12\} and \{34\} each cover but a small portion of the 
model universe, and corresponds geometrically to the polar caps. 
RelSize(1,2) small means physically that cluster \{12\} is but a speck in the firmament, and corresponds 
geometrically to a belt around the `Bangladeshi' meridian. 
RelSize(1,2) large means physically that cluster \{12\} engulfs the rest of the model universe, and 
corresponds geometrically to an antipodal pair of caps around each of the intersections of the equator 
and the Greenwich meridian.

A quantifier of the contents inhomogeneity between the two clusters is $\Phi$, which is related to the 
ratio of the size of \{34\} to that of \{12\}'s by (\ref{Var}).  
N.B. that the last 3 paragraphs refer, more concretely, to $\sqrt{\mbox{mass}} \times$ length, so that 
large mass hierarchies can distort intuitive notions of `actual size'.

\subsection{Rotor and planar mechanics analogies for 4-stop metroland}

By inspection of the kinetic term, there are clear analogies between this 4-stop metroland problem and 
well-known rotor and planar problems in ordinary mechanics. 
For the first analogy, 
\beq
\mbox{arctan}
\big(
\sqrt{ \mbox{RelSize}(12)^2 + \mbox{RelSize}(34)^2}    /    \mbox{RelSize(12,34)}    
\big)
= \Theta \longleftrightarrow \theta = \mbox{ (azimuthal coordinate of the axis in space) }
\label{ana3}
\eeq
\beq
\mbox{arctan}
\left(
\mbox{RelSize}(34)/\mbox{RelSize}(12)
\right) = \Phi \longleftrightarrow \phi = (\mbox{polar coordinate of the axis in space}) \mbox{ } ,
\label{ana2}
\eeq
\beq
1 \leftrightarrow I_{\sr\so\st\so\sr} \mbox{ } \mbox{ } \mbox{( moment of inertia of the rotor )} 
\mbox{ } \mbox{ } .  
\label{ana4} 
\eeq
For the second analogy, transform $\Theta$ to the radial stereographic coordinate ${\cal R} = 
\mbox{tan$\frac{\Theta}{2}$}$ and pass to the `tilded' banal representation.   
One then has the flat plane polar coordinates kinetic term, so 
\beq
\sqrt{    \{1 - \mbox{RelSize}(12,34)\}/\{1 + \mbox{RelSize}(12,34)\}    } = {\cal R} 
\longleftrightarrow r = \mbox{(radial coordinate of test particle)}  \mbox{ } , 
\label{an1}
\eeq
\beq
\mbox{arctan}
\left(
\mbox{RelSize}(34)/\mbox{RelSize}(12)
\right) = \Phi \longleftrightarrow \phi = (\mbox{polar coordinate of test particle}) \mbox{ } ,
\label{an2}
\eeq
\beq
1 \leftrightarrow m =(\mbox{test particle mass}) \mbox{ } .
\label{an3}
\eeq
These analogies will be furtherly fruitful in analyzing 4-stop metroland's equations of motion 
and conserved quantities in the next 2 subsections, as well as when further specifics about the potential 
are brought in (see Sec 3.6).

\subsection{Equations of motion for 4-stop metroland}

The equations of motion are 
\beq
\Theta^{**} - \mbox{sin}\,\Theta\,\mbox{cos}\,\Theta\,{\Phi^*}^2 = -\fV_{,\Theta}  \mbox{ } , \mbox{ } 
\{\mbox{sin}^2\Theta\,\Phi^*\}^* = -\fV_{,\Phi}  \mbox{ } .
\eeq
(The star is derivative with respect to the relational approach's emergent time $t$: 
$* \equiv \d /\d t \equiv \sqrt{\{\fE - \fV\}/{\fT}} \mbox{ } \dot{\mbox{}}$, for which the equations of 
motion simplify.  
This is readily deduced to banal-transform as $* \longrightarrow \Omega^{-2}*$ \cite{Banal}.]   


$\fV$ is independent of $\lambda$ itself and so one of these can be replaced by the `energy relation' 
(a first integral):
\beq
\{{\Theta^*}^2 + \mbox{sin}^2\Theta\,{\Phi^*}^2\}/2 + \fV(\Theta, \Phi) = \fE \mbox{ } , \mbox{ constant .}
\eeq
If the potential is additionally $\Phi$-independent (which we term `special' and whose planar mechanics 
analogue is termed central), then the $\Phi$-Euler--Lagrange equation above gives another first integral,  
\beq
\mbox{sin}^2\Theta\, \Phi^* = {\cal D} \mbox{ } .  
\label{arva}
\eeq
For both of the analogies above, the correpsonding SO(2) or SO(3) related constant of the motion has the 
physical meaning of an angular momentum; for its interpretation in the present context, however, see the 
next subsection.  
In the special

\noindent case, one can now furthermore combine the last 2 equations in two ways. 
Firstly,  
\beq
\fE = {\Theta^*}^2/2 + {\cal D}^2/2\mbox{sin}^2\Theta + \fV(\Theta) \equiv 
{\Theta^*}^2/2 + \fV_{\se\sf\sf} \mbox{ }
\eeq
which in the planar central problem amounts to modification of the potential by a centrifugal barrier, 
while, in the more directly analogous rotor problem, amounts to placing a centrifugal barrier at each 
pole.  
In our problem, it takes the latter `bipolar barrier' form.
Secondly, (for ${\cal D} \neq 0$)
$
\Theta_{,\Phi}\mbox{}^2/2 = \mbox{sin}^4\Theta\{\fE - \fV_{\se\sf\sf}\}/{\cal D}^2  
$.
Both of these straightforwardly give quadratures relating $\Theta$ to, respectively, $t$ 
(orbit traversal rate) and $\Phi$ (shape of the orbit, the ${\cal D} = 0$ case giving a $\Phi$ = 
constant 1-$d$ motion without any double barrier).

If $\fV$ is also $\Theta$-independent and thus constant, we get 3 ${\cal D}$-quantities from freedom 
to pick whichever axis to have a conserved $\Phi$ about.  
We call this constant-potential case the `very special case' (the counterpart of which in the second 
analogy is the rigid rotor).

\subsection{Further discussion of 4-stop metroland's conserved quantitites}

One interesting issue in RPM's is what these angular momentum-like quantities are physically 
Triangleland is spatially 2-$d$ and as such affords a notion of angular momentum; its 
conserved quantity ${\cal J}$ turns out to be the {\sl relative} angular momentum 
between its subsystems.  
But the present paper's 4-stop metroland problem, however, is spatially 1-$d$, so no angular momentum in 
space (relative or otherwise) is possible.  
What then is the meaning of the conserved quantity ${\cal D}$ in terms of the $n^i$ or RelSize variables?
\beq
{\cal D} = n_x{n_y}^* - n_y{n_x}^* \mbox{ } , 
\label{J}
\eeq
which is the `3-component of an angular momentum in the Euclidean relational 
configuration space $R(4, 1) = \mathbb{R}^3$'.  
Moreover, using $\ttD_i$ for individual/partial dilations $R^iP_i$ (no sum) 
\beq
{\cal D} = \ttD_2n_x/n_y - \ttD_1n_y/n_x
\eeq 
so that ${\cal D}$ is a (weighted) {\it relative dilational quantity} corresponding to a particular 
exchange of dilational momentum between the \{12\} and \{34\} clusters.

In the very special case, there are 3 ${\cal D}$ conserved quantities forming a vector in the 
`surrounding' Euclidean relational configuration space $\mathbb{R}^3$, of which the above ${\cal D}$ is 
the 3-component: 
\beq
{\cal D}_i = \epsilon_{ijk}n^jn^{k\,*} = \ttD_kn^j/n^k  - \ttD_jn^k/n^j 
\eeq
where $i$, $j$, $k$ are a cycle of 1, 2, 3.  

All in all, the less special a problem is, the more types of relative dilational momentum exchanges it 
has.

That we get `angular momentum mathematics', we explain as follows. 
The body of mathematics habitually associated with angular momentum can actually be associated more 
generally (in terms of what physics it covers, not what mathematics it is, as further explained in 
Appendix A) with rational (i.e. `ratio-based') quantities rather than just with angular ones (which are 
a subset thereof).\footnote{Smith  
\cite{Smith} pointed out this generalization but not, as far as we are aware of, its rational 
interpretation.}
Therefore `rational momentum mathematics' would appear to be a more widely appropriate term, covering 
both angular momentum and dilational momentum as subcases.  
The objects in question continue to possess antisymmetry in this more general setting since this 
derives from differentiating a (function of a) ratio by the (chain rule and) quotient rule: 
$\{f(y/x)\}^{*} = f^{\prime}\{xy^* - yx^*\}/x^2$ and thus occurs irrespective of whether that ratio 
admits an interpretation as an angle in physical space.

\subsection{Passage to 4-stop metroland's Hamiltonian}

The conjugate momenta are then 
\beq
p_{\Theta} = \Theta^* \mbox{ } ,  \mbox{ } 
p_{\Phi} = \mbox{sin}^2\Theta\,\Phi^* = {\cal D} \mbox{ } .  
\eeq
[Also, ${\cal D}_i = {\epsilon_{ij}}^k\iota^jp_k$ for $p_k$ the momentum conjugate to $\iota^k$.]
The momenta obey a quadratic constraint 
\beq
\ttQ \equiv 
{p_{\Theta}}^2/2 + {{p_{\Phi}}^{2}}/2{\mbox{sin}^2\Theta} 
+ \fV(\Theta, \Phi) = \fE \mbox{ } ,
\eeq
the middle expression of which also serves as the classical Hamiltonian $\fH$ for the system.

\subsection{Harmonic oscillator like potentials for 4-stop metroland}

With eventual timeless records and structure formation goals in mind, we intend to 
follow a particular clustering -- the \{12,34\} one -- using a particular permutation of H-coordinates 
which is physically picked out by considering not the most 
general array of 6 springs between the 
particles but rather the following. 
We take the mechanical picture in Jacobi 
coordinates as primary and consider springs within each of the 
\{12\} and \{34\} clusters and between the centres of mass

\noindent of the two clusters (reinterpretable if one so 
wishes as a superposition of inter-particle springs).  
Then the potential is 
$$
\fV = \sumi3 K_i \{n^i\}^2/2 = \{K_1^2\,\mbox{sin}^2\Theta\,\mbox{cos}^2\Phi 
                                                   + K_2^2\,\mbox{sin}^2\Theta\,\mbox{cos}^2\Phi  
                                                   + K_3^2\,\mbox{cos}^2\Theta\}/2 = 
$$
\beq 
A + B\mbox{cos}2\Theta + C \mbox{sin}^2\Theta\mbox{cos} 2\Phi = 
a + bY_{2,0}(\Theta) + cY_{2, 2\sc}(\Theta, \Phi) 
\label{39}
\eeq
for $K_i = H_i/\mu_i$ where $H_i$ play the role of Jacobi--Hooke coefficients, and

\noindent
\beq
A = K_3/4 + \{K_1 + K_2\}/{8} \mbox{ } , \mbox{ } 
B = K_3/4 - \{K_1 + K_2\}/{8} \mbox{ } , \mbox{ } 
C = \{K_1 - K_2\}/{4} \mbox{ } ,  
\eeq
the $Y$'s are spherical harmonics (c and s subscripts thereon standing for cosine and sine $\Phi$-parts) 
and the precise form of the constants $a$, $b$, $c$ is not required for this paper.  
This potential has as a `very special' case $B = C = 0$, for which the potential is constant, and 
the `special case' $C = 0$ for which the dynamics is separable (which is sketched in Fig 3).   
In terms of the $K_i$, the special case corresponds to $K_1 = K_2$, i.e. that each cluster has the same 
`constitution': the same Jacobi--Hooke coefficient per Jacobi cluster mass, which is a kind of `homogeneity 
requirement' on the `structure formation' in the cosmological analogy.  
The very special case then corresponds to $K_1 = K_2 = K_3$, for which high-symmetry situation the various 
potentials can balance out to produce the constant. 
Additionally the $B << A$ perturbative regime about the very special case signifies 
$K_1 + K_2 << K_3$ so the inter-cluster spring is a lot stronger than the intra-cluster springs, which 
in some ways is analogous to scalefactor dominance over inhomogeneous dynamics in cosmology.  
On the other hand, the $C << A$ regime corresponds to either or both of the conditions 
$K_1 + K_2 << K_3$, $K_1 \approx K_2$ the latter of which signifies high contents homogeneity. 
The multiplicity of forms of writing the potential above is useful to bear in mind in searching for 
mathematical analogues for the present problem in e.g. the molecular physics literature (c.f. Sec 3.6).

If one started instead with springs between all 6 pairs of particles, one would obtain 
$\fV_6 = \fV + \fV^{\prime}$ for 
\beq
\fV^{\prime} = 
\sumi3 L_i n^j n^k =  D\mbox{sin}^2\Theta\,\mbox{sin}\,2\Phi + 
      E\mbox{sin}\,2\Theta\,\mbox{cos}\,\Phi + F\mbox{sin}\,2\Theta\,\mbox{sin}\,\Phi  = 
dY_{2,2\sss}(\Theta, \Phi) + eY_{2,1\sc}(\Theta, \Phi) + fY_{2,1\sss}(\Theta, \Phi)   
\label{3More}
\eeq
where $i$, $j$, $k$ are a cycle so that the first form of $\fV_6$ is the most general homogeneous 
quadratic polynomial in the $n^i$, $D = L_3/2$, $E = L_2/2$ and $F =L_1/2$, and the detailed form 
of the constants $d$, $e$ and $f$ are not needed for this paper.  
On the face of it, this is a more general problem than the preceding paragraph's, with 
three further nonseparable terms.  
However, there is a sense in which these three terms can be made to go away, c.f. Sec 2.12.   
One can imagine whichever of these problems' potentials as a superposition of familiar `orbital shaped' 
lumps, though such a superposition will of course in general alter the number, size and position of 
peaks and valleys according to what coefficients each harmonic contribution has.  
Contrast with the triangleland model is also interesting at this point -- there $Y_{0,0}$ and just two 
of the first-order spherical harmonics arose. 

\mbox{ }

The equations of motion for this potential are 
\beq
\Theta^{**} - \mbox{sin}\,\Theta\,\mbox{cos}\,\Theta\,{\Phi^*}^2 = 
\mbox{sin}\,2\Theta\,\{2B - C\,\mbox{cos}\,2\Phi - D\,\mbox{sin}\,2\Phi\} + 
2\,\mbox{cos}\,2\Theta\,\{E\,\mbox{cos}\,\Phi + F\,\mbox{sin}\,\Phi\} \mbox{ } ,
\eeq
\beq
\{\mbox{sin}^2\Theta\, \Phi^*\}^* = 
2\,\mbox{sin}^2\Theta\,\{C\,\mbox{sin}\, 2\Phi - D\,\mbox{cos}\,2\Phi\} + 
\mbox{sin}\,2\Theta\,\{E\,\mbox{sin}\,\Phi - F\,\mbox{cos}\,\Phi\} \mbox{ } ,  
\eeq
one of which can be replaced by the `energy' first integral
\beq
\{{\Theta^*}^2 + \mbox{sin}^2\Theta\,{\Phi^*}^2\}/2 + 
A + B\,\mbox{cos}\,2\Theta + C\,\mbox{sin}^2\Theta\,\mbox{cos}\,2\Phi + 
D\,\mbox{sin}^2\Theta\,\mbox{sin}\,2\Phi + E\,\mbox{sin}\,2\Theta\,\mbox{cos}\,\Phi + 
F\,\mbox{sin}\,2\Theta\,\mbox{sin}\,\Phi = \fE \mbox{ } .  
\eeq
Then if $C = D = E = F = 0$, one has a special potential, so the $\Phi$ Euler-Lagrange equation 
gives another first integral (\ref{arva}) and the subsequent quadrature for the shape of the orbit is 
\beq
\Phi - \Phi_0 = \pm{\cal D}\int\d\Theta
\left/
\mbox{sin}\,\Theta\sqrt{2\{\fE - A - B\,\mbox{cos}\,2\Theta\}\mbox{sin}^2\Theta - {\cal D}^2} \mbox{ } . 
\right.
\eeq


{            \begin{figure}[ht]
\centering
\includegraphics[width=0.4\textwidth]{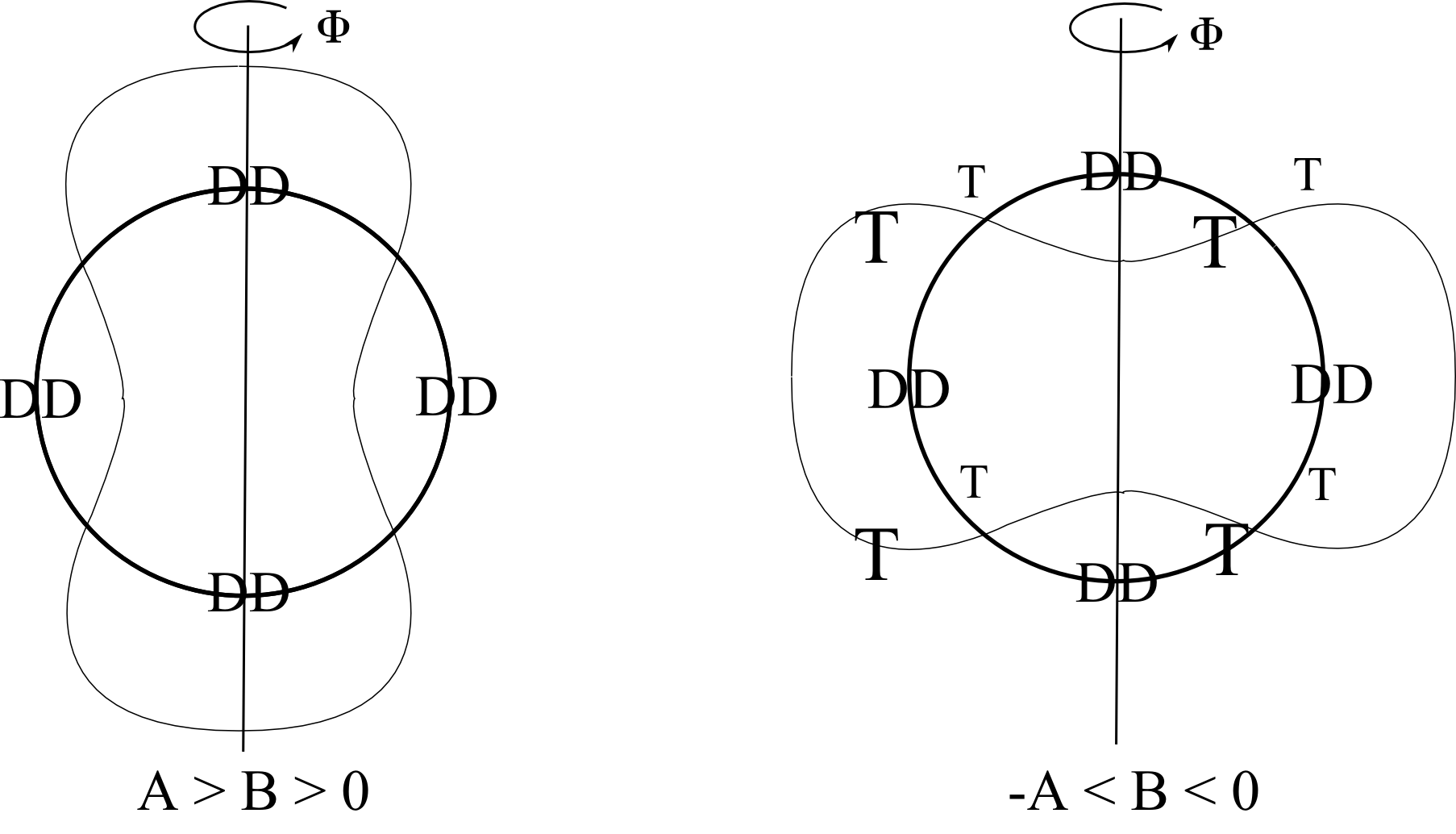}
\caption[Text der im Bilderverzeichnis auftaucht]{        \footnotesize{    We sketch $\fV$ over the 
sphere for the mechanically significant cases a) $A > B > 0$ and b) $-A < B < 0$.  
The first has barriers at the poles and a well around the equator, while the second has wells 
at the pole and a barrier around the equator.  
Each is a surface of revolution of the curve provided, the first case being a peanut or 
ellipsoid-like surfaces and the second case being a concave or convex wheel.  
In each case, 
considering $\fV_{\se\sf\sf}$ for ${\cal D} \neq 0$ adds a spike at each pole  
(this now means that for ${\cal D} \neq 0$ both islands cannot simultaneously collapse to their 
generally distinct centre of mass points).
Finally, note that our potential is axisymmetric and reflectible about its equator so its symmetry 
group is $\mathbb{D}_{\infty} \times \mathbb{Z}_2$ ($\mathbb{D}$ denotes dihaedral).  
If this is aligned with a DD axis of the physical interpretation, the overall problem retains a 
$\mathbb{D}_4 \times \mathbb{Z}_2$ symmetry group, of order 16.  
    }        }
\label{Fig3}\end{figure}           }

One simple consideration here is small and large regimes for the special case.  
More precisely, these are near-North Pole and near-South Pole regimes in $\Theta$ but become large and 
small regimes in terms of ${\cal R} = \mbox{tan$\frac{\Theta}{2}$}$.  
For this (including changing to the tilded banal representation), and using `shifted energy' 
$\fE^{\prime} \equiv \fE - A - B$
\beq
\fW \equiv \widetilde{\fE} - \widetilde{\fV} = 4\fE^{\prime}/{\{1 + {\cal R}^2\}^2} + 
{32B{\cal R}^2}/{\{1 + {\cal R}^2\}^4} \mbox{ } . 
\label{r}
\eeq
Then the near-North Pole regime (${\cal R} << 1$) maps to the problem with flat polar kinetic term and  
\beq
\fW = 4\fE^{\prime} + 8\{4B - \fE^{\prime}\}{\cal R}^2 
\label{oc}
\eeq
up to $O({\cal R}^4)$.  
This has the mathematics of a 2-$d$ isotropic harmonic oscillator,  
\beq
\fW = {\cal E} - \omega^2{\cal R}^2/2 \mbox{ } ,
\label{qus} 
\eeq
provided that the `classical frequency' (for us with units of $I$/time) $\omega \mbox{ } < 0$ (else it 
would be a constant potential problem or an upside-down harmonic oscillator problem), alongside 
${\cal E} > 0$ to stand a chance of then meeting classical energy requirements.  
Writing ${\cal E}$ and $\omega^2$ out by comparing the previous two equations, these inequalities 
signify that $2\fE > K_3$ and $2\fE > K_3 + 2\{K_3 - K_1\}$, the latter being more stringent if 
$K_3 > K_1$ (`stronger inter-cluster binding') and less stringent if $K_3 \leq K_1$ 
(`weaker inter-cluster binding').
One can also deduce from the first of these and $K_3 \geq 0$ (spring) that $\fE > 0$.

Next, note that the near-South Pole regime (${\cal R} << 1$) maps to the problem with flat polar kinetic 
term and  
\beq
\fW = 4\fE^{\prime}/{{\cal R}^4} + {8\{4B - \fE^{\prime}\}}/{{\cal R}^6} 
\label{w}
\eeq
up to  $O({1}/{\cal R}^8)$. 
Moreover, ${\cal U} = 1/{\cal R}$ maps the large case's (\ref{r}) to the small case's (\ref{w}), 
so this is also an isotropic harmonic oscillator -- in $({\cal U}$, $\Phi$) coordinates and with the 
same ${\cal E}$ and $\omega$ as above.  
One of us had previously observed a `large--small' duality of this sort in triangleland \cite{08I}. 
It halves the required solving to understand $\Theta$ $\approx 0$ and $\approx \pi$. 
Another lesson learnt from the triangleland study is that we know that study of {\sl second} 
approximations is considerably more profitable than that of first approximations, so we pass straight to them.  
Note that, for our subsequent QM study, we want the isotropic harmonic oscillator rather than cases 
corresponding to other values of the parameters ${\cal E}$ and $-\omega^2$ (e.g. the upside-down isotropic 
harmonic oscillator).

\subsection{Classical solutions for ${\cal D} = 0$}

$0 = {\cal D} = \mbox{sin}^2\Theta\,\Phi^*$ so either $\sin\Theta = 0$ and one is stuck on a pole or 
$\Phi$ is constant.
In terms of the $n^i$, this translates to 
$n_y = kn_x$ for $k$ constant, so motion is restricted to lying on a diameter.  
In the case of $\underline{\cal D} = 0$ (which corresponds to constant potential case with 
${\cal D} = 0$), one likewise obtains $n_y = kn_x = ln_z$ for $l$ also constant, but 
(\ref{New19}) holds too, so all $n^i$ take fixed values and motion is restricted to a point.  
Being purely 1-$d$ motions or 0-d motions, this subsection's solutions' simpleness renders them of limited 
interest. 
1-$d$ motions include 1) going up and down the 1-axis, corresponding to cluster \{34\} always being 
collapsed while cluster \{12\} varies in size including going through zero size at the origin and two 
triple collisions in which each of particles 1 and 2 coincide with the collapsed cluster. 
2) The \{12\} $\leftrightarrow$ \{34\} of this going up and down the 2-axis.  
3) Going up and down an $n_y = \pm n_x$ line, corresponding to the clusters always being of the 
same size (contents homogeneity) but that size varying from zero (\{12,34\} DD collision, i.e. \{12\} 
collapsing to a point and also \{34\} collapsing to a point) to maximal [in which the two clusters are 
superposed into the \{13,24\} or \{14,23\} DD collisions].

\subsection{Classical solution in the very special case}

For ${\cal D} \neq 0$, the very special case is solved by the geodesics on the shape space sphere, 
\beq
\mbox{cos}(\Phi - \Phi_0) = \kappa\,\mbox{cot}\,\Theta \mbox{ } 
\eeq  
for $\kappa = {\cal D}/\sqrt{2\{\fE - A\} - {\cal D}^2}$, constant.  
Then in terms of the $n^i$ (or RelSize variables), (\ref{useful} i--iv) gives 
\beq
\kappa n_z = n_x\,\mbox{cos}\,\Phi_0 + n_y\,\mbox{sin}\,\Phi_0 \mbox{ } ,
\label{plane}
\eeq
i.e. restriction to a plane through the origin, with arbitrary normal (cos$\,\Phi_0$, sin$\,\Phi_0$, 
$-\kappa$).      
But also $\sumi3 \{n^i\}^2 = 1$, so we are restricted to the intersection of the sphere and the arbitrary 
plane through its centre, which is clearly another well-known way of describing the great circles as 
circles within $\mathbb{R}^3$.

The disc in the equatorial plane is particularly useful for considering the mechanics of the problem 
with clusters \{12\} and \{34\} picked out by our choice of Jacobi H-coordintes.  
Eliminating $n_z$ projects an ellipse onto this disc,
\beq
\kappa^2 = \{\kappa^2 + \mbox{cos}^2\Phi_0\}{n_x}^2 + 2\,\mbox{cos}\,\Phi_0\,\mbox{sin}\,\Phi_0n_xn_y 
+ \{\kappa^2 + \mbox{sin}^2\Phi_0\}{n_y}^2 \mbox{ } ,  
\eeq
centred on the origin with its principal axes in general not aligned with the coordinates.  
E.g. for $\Phi_0 = 0$, the ellipse is
\beq
\big\{{\mbox{RelSize(1,2)}}/{\{1 + \kappa^{-2}\}^{-1/2}}\big\}\mbox{}^2 + {\mbox{RelSize(3,4)}}^2 = 1
\mbox{ } ,
\eeq
which has major axis in the RelSize(3,4) = $n_y$ direction and minor axis in the RelSize(1,2) = $n_x$ 
direction, while the value of RelSize(12,34) = $n_z$ around the actual curve can then be read off 
(\ref{plane}) to be $n_z = n_x/\kappa$.  
With reference to the first subfigure in Figure 2a), as ${\cal D} \longrightarrow \infty$, the dynamical 
trajectory is the equator, corresponding to maximally-merged configurations including four DD collisions.  
For ${\cal D}$ small, the motion approximately goes up and down a meridian, e.g. forming a basic unit of 
a narrow cycle from the polar DD to slightly around the T on the Greenwich meridian (reflections of) 
which is repeated various times to form the whole trajectory. 
[The actual limiting on-axis motion 
${\cal D} = 0$ is excluded from this subsection's working but already considered in the preceding one.]


Other $\Phi_0$ straightforwardly correspond to rotated ellipses.  However the mechanical meaning of 
these differs. 
E.g. about $\pi/2$ clusters \{12\} and \{34\} are interchanged, while about $\pm \pi/4$ also has distinct 
sharp physical significance.  
Throughout, note the periodicity of the motion (already clear in the spherical model as the great circles 
are closed curves).  
The tessellation lines are great circles, projecting to the disc rim, the axes, the lines at $\pi/4$ 
to the axes and ellipses with principal directions aligned with the preceding.

\subsection{Approximate classical solutions in the special case}

At the level of the sphere and using the second approximation, we can transcribe the solution from 
\cite{08I} to be, with ${\cal D}$ for ${\cal J}$ and with our ${\cal E}$ and $\omega^2$ in place of 
that paper's $Q_0$ and $Q_2$, and defining $f_0 = 2{\cal E}/{\cal D}^2$, $f_2 = \omega^2/{\cal D}^2$ and 
$g = \sqrt{{f_0}^2 - f_2}$, 
\beq
\sqrt{f_0 + g\,\mbox{cos}(2\{\Phi - \Phi_0\})} = {1}/{{\cal R}} = \mbox{cot$\frac{\Theta}{2}$} \mbox{ } .  
\label{ellipse}
\eeq
In terms of ${\cal R}$ these are ellipses centred on the origin (including the bounding case of circles 
but excluding the other bounded case of pairs of straight lines (Fig 4).

{            \begin{figure}[ht]
\centering
\includegraphics[width=0.5\textwidth]{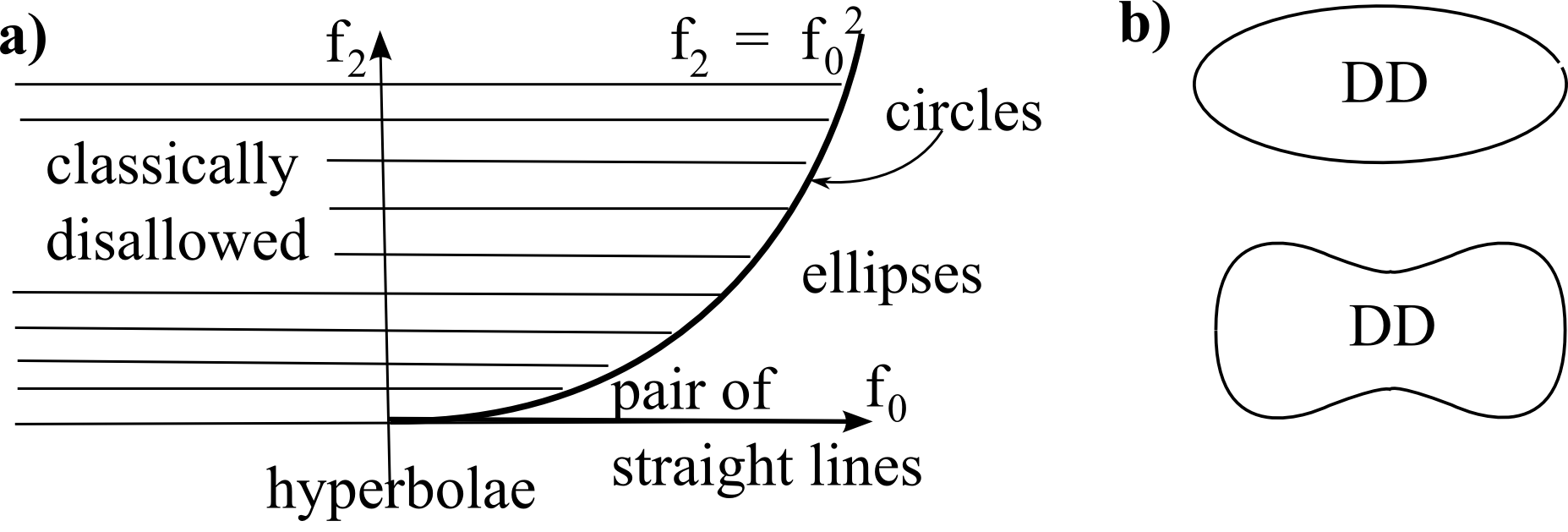}
\caption[Text der im Bilderverzeichnis auftaucht]{        \footnotesize{    a) For the small 
approximation in (${\cal R}$, $\Phi$) coordinates, the desired parameter space is the indicated wedge 
populated by ellipses; on the bounding parabola, we get circles, coinciding with the edge of the $n_x$, 
$n_y$ disc for $f_0 = 1 = f_2$ and becoming smaller in either direction. 

b) For the large approximation in (${\cal W}$, $\Phi$) coordinates, the parameter space is likewise; 
if one converts to (${\cal R}$, $\Phi$) coordinates, however, the wedge is then populated by 
ellipse-like curves and peanut-like curves.
}        }
\label{Fig4}
\end{figure}            }


\noindent
N.B. we now have a third inequality on ${\cal E}$ and $\omega$:   
$
4{\cal E}^2 \geq {\cal D}^2\omega^2, 
$
that replaces ${\cal E} > 0$ as it is more stringent.
Thus in terms of $\fE$ and the $K_i$ we get our allowed wedge of parameter space to be 
$
2\{2\fE - K_3\}^2/{\cal D}^2 \geq \{2\fE - K_3\} + 2\{K_1 - K_3\} > 0.   
$
Saturation of this corresponds to circular trajectories.   
For such circles to exist, the discriminant gives the condition $\{{\cal D}/4\}^2 \geq K_3 - K_1$, so 
that the relative dilational quantity is bounded from below by the amount by which the 
inter-cluster spring dominates.

Then by (\ref{useful} v-vi) and (\ref{Rdef}), (\ref{ellipse}) becomes  
\beq
\sqrt{        f_0 + g      \{    \{{n_x}^2 - {n_y}^2\}\,\mbox{cos}\,2\Phi_0 + 
                                 2 n_xn_y \,\mbox{sin}\,2\Phi_0    \}/
                           \{    {n_x}^2 + {n_y}^2                 \}         } = 
\sqrt{\{1 + n_z\}/\{1 - n_z\}} \mbox{ } ,
\eeq
so solving for $n_z$ and applying the on-sphere condition for $\Phi_0 = 0$, say, gives 
\beq
\sqrt{    1 - {n_x}^2 - {n_y}^2    } = n_z = 
\frac{    \{f_0 + g - 1\}{n_x}^2 + \{f_0 - g - 1\}{n_y}^2    }
     {    \{f_0 + g + 1\}{n_x}^2 + \{f_0 - g + 1\}{n_y}^2    } \mbox{ } .  
\eeq
Then one can write down a curve in terms of two independent variables such as RelSize(1,2) and 
RelSize(3,4): either RelSize(1,2) + RelSize(3,4) = 0 (so both are 0 because they are positive and so 
both clusters have collapsed) or
\beq
\big\{\{f_0 + g + 1\}\mbox{RelSize(1,2)}^2 + \{f_0 - g + 1\}\mbox{RelSize(3,4)}^2\big\}\mbox{}^2 = 
4\{\{f_0 + g\}\mbox{RelSize(1,2)}^2 + \{f_0 - g\}\mbox{RelSize}(3,4)^2\} \mbox{ } .
\eeq

The large regime then has
\beq
\sqrt{f_0 + g\,\mbox{cos}(2\{\Phi - \Phi_0\})} = {1}/{\cal W} = {\cal R} = \mbox{tan$\frac{\Theta}{2}$} 
\mbox{ } ,
\eeq
which is, for the cases of interest, an ellipse-like or peanut-like curve (see \cite{08I} and Fig 4b). 
Applying this paper's interpretation in terms of $n^i$ or RelSize variables, the {\sl same} answer as 
for the small regime arises again.    
This conclusion just reflects the potential imposed having an antipodal symmetry, which physically 
translates to shapes and their mirror images behaving in the same fashion.  


Finally, note that this approximate problem has a part-hidden SO(3) symmetry (such are well-known for 
harmonic oscillators).     
In the present context, however, its objects take the form  
\beq
{\cal H}_1 = \omega n_xn_y + \ttD_1\ttD_2/\omega n_xn_y 
\mbox{ } \mbox{ } , \mbox{ } \mbox{ }
{\cal H}_2 = {\cal D} = \ttD_2n_x/n_y - \ttD_1n_y/n_x  
\mbox{ } \mbox{ } , \mbox{ } \mbox{ }
{\cal H}_3 = \omega\{{n_x}^2 - {n_y}^2\}/2 + 
{\ttD_1}^2/2\omega{n_x}^2 - {\ttD_2}^2/2\omega {n_y}^2
\mbox{ } . 
\eeq 
Thus its unhidden part is ${\cal H}_2 = {\cal D}$ which has relative dilational momentum significance, 
while its remaining hidden parts are mixed shape and dilational objects.  
[This is used as an example in the discussion in Appendix A.]

\subsection{Discussion of `more general combinations of springs'}

The $L_i$ terms (or, equivalently, $D$, $E$ and 
$F$ terms) can be dropped in the sense that one can pass to normal coordinates for which the symmetric 
matrix of Jacobi--Hooke coefficients has been diagonalized. 
Unlike in triangleland, however, this does not send one to the special case -- the $C$-term survives 
and so requires addressing separately (e.g. perturbatively).  
The elimination thus of $D$, $E$ and $F$ terms is also subject to the mechanical interpretation of the 
normal-coordinate problem being more difficult algebraically than for $D$ = $E$ = $F$ = 0, so that  
it is conceivable that one might prefer to retain this simpler interpretation and treat $D$, $E$ and $F$ 
perturbatively.

\section{4-stop metroland at the quantum level}

Kinematical quantization \cite{Isham84} for this problem \cite{08II} involves three objects $u^i$ 
whose squares add up to 1 (which in the present case we identify with the Cartesian unit vectors $n^i$) 
and three SO(3) objects (which in the present case are the ${\cal D}_i$).  
We then consider the time-independent Schr\"{o}dinger equation \cite{08II}
\beq
\{\mbox{sin}\,\Theta\}^{-1}
\big\{\mbox{sin}\,\Theta\,\Psi_{,\Theta} 
\big\},_{\Theta} 
+  \{\mbox{sin}\Theta\}^{-2}\Psi_{,\Phi\Phi} = 
\{{\cal A} + {\cal B}\,\mbox{cos}\,2\Theta + {\cal C}\mbox{sin}^2\Theta\,\mbox{cos}\,2\Phi\}\Psi \mbox{ } , 
\label{62}
\eeq
where ${\cal A} = 2\{{A} - {\fE}\}/\hbar^2$, ${\cal B} = 2{B}/\hbar^2$, ${\cal C} = 2{C}/\hbar^2$ are 
dimensionless constants.   
Note that the above equation is separable for $0 = {\cal C}$ i.e. $0 =  C$ i.e. $K_1 = K_2$;   
most of our work is for this case.

\subsection{Explanation of the choice of operator ordering}

We choose an ordering that is coordinatization invariant on configuration space \cite{DeWitt57}, i.e. a 
member of the family $D^2 - \xi\mbox{Ric}(M)$ [c.f. (\ref{WDE}, \ref{RPMWDE})] where $D^2$ and 
$\mbox{Ric}(M)$ are the Laplacian and the Ricci scalar corresponding to the kinetic metric 
$M$ on configuration space.   
Moreover, following from the appropriateness of relational actions for whole-universe physics, 
observing that these have banal conformal invariance as a simple and natural feature and then asking 
for this to hold at the quantum level in the whole-universe context (i.e. in quantum cosmology or toy 
models thereof), among the preceding family of orderings we are uniquely led to the {\sl conformal 
ordering}, for which $\xi =\{ k - 2\}/4\{k - 1\}$, where $k$ is the configuration space dimension.  
(See \cite{Banal} for more on this motivation for conformal ordering, previous motivation on different premises 
for it being in e.g. \cite{Magic, Oporder}.)
Moreover, presently we are in configuration space dimension 2, for which $\xi = 0$ so our operator 
ordering choice is, in this case the same as the Laplacian ordering (itself advocated in e.g. \cite{Lap}, 
while \cite{Page91LoukoBarvin} also considered 2-$d$ configuration spaces so the Laplacian--conformal 
ordering coincidence also applies).

\subsection{Solution in very special case}

The ${\cal C} = 0$ case of Eq (\ref{62}) separates to simple harmonic motion and the $\Theta$ equation 
\beq
\{\mbox{sin}\,\Theta\}^{-1}
\{\mbox{sin}\,\Theta\,\Psi_{,\Theta} 
\}_{,\Theta} 
- \{\mbox{sin}\,\Theta\}^{-2}\mm^2\Psi  = 
{\cal A}\Psi + {\cal B}\,\mbox{cos}\,2\Theta\,\Psi \mbox{ } , 
\label{chi}
\eeq
If ${\cal B}$ = 0 as well -- our very special problem --, then from Sec 2.4 this has similar mathematics 
to ordinary QM's central potential problem, in which the quantum Hamiltonian $\hat{\fH}$, total angular 
momentum $\widehat{\ttL}_{\sT\so\st\sa\sll} = \sum_{\alpha = 1}^3\widehat{\ttL}_\alpha\mbox{}^2$ and 
magnetic/axial/projected angular momentum $\hat{\ttL}_3$ form a complete set of commuting 
operators and as such share eigenvalues and eigenfunctions.  
In fact (also Sec 2.4) our very special problem is mathematically the same as the rigid rotor, for 
which $\widehat{\fH}$ {\it is} $\ttL_{\sT\so\st\sa\sll}$ up to multiplicative and additive constants, 
so, effectively one has a complete set of two commuting operators, whose eigenvalues and eigenfunctions 
are the well-known spherical harmonics and, moreover also occur as a separated-out part of the 
corresponding scaled relational particle model problem.   
However, our `rigid rotor' is in configuration space rather than in space and with total relative 
dilational momentum $\widehat{{\cal D}}_{\sT\so\st\sa\sll} = \sum_{i=1}^3\widehat{{\cal D}_i}\mbox{}^2$ in 
place of total angular momentum and projected relative dilational momentum $\widehat{{\cal D}}_3$ in 
place of axial angular momentum.  
These then have eigenvalues $\hbar^2\mD\{\mD + 1\}$ and $\hbar\d$ respectively, so we term D and d 
respectively the 
{\it total} and {\it projected relative dilational quantum numbers} (which are analogous to the ordinary 
central force problem/rigid rotor's total and axial/magnetic angular momentum quantum numbers).

Our very special problem's time-independent Schr\"{o}dinger equation separates into simple harmonic 
motion and the associated Legendre equation (in $X = \mbox{cos}\,\Theta$) i.e. the spherical harmonics 
equations, 
Thus its solutions are 
\beq
\Psi_{\sD\sd}(\Theta, \Phi) \propto Y_{\sD\sd}(\Theta, \Phi) \propto 
\mP_{\sD}^{\sd}(\mbox{cos}\,\Theta)\mbox{exp}(\pm i\d\Phi) \mbox{ } 
\eeq
for $\mP_{\sD}^{\sd}(X)$ the associated Legendre functions of $X$, D $\in \mathbb{N}_0$ and d such that 
$|\d| \leq \mD$.  
Also, $\mD\{\mD + 1\} = - {\cal A}$, which, interpreted in terms of the original quantities of the 
problem, is the condition 
\beq
\fE^{\prime} = \fE - K_3/2 = \hbar^2\mD\{\mD + 1\}/2
\eeq 
on the model universe's `energy' and inter-cluster effective spring in order to have any quantum solutions 
($\fE$ is {\sl fixed} as this is a whole-universe model so there is nothing external from which it could 
gain or lose energy).  
If this is the case, there are then 2D + 1 solutions labelled by d (we can see the preceding sentence 
cuts down on a given system's solution space, though the more usual larger solution space still exists 
in the `multiverse' sense \cite{+Tri}).

Furthermore, using a basis with sines and cosines instead of positive and negative exponentials, 
\beq
\Psi_{\sD_{\mbox{}_{\cal N}}}(n^i) \propto {\cal N}(n^i) \mbox{ } . 
\eeq
Here, the D-label runs over the orbital types ($s$ for D = 0, $p$ for D = 1, $d$ for D = 2 ...) and 
${\cal N}$ is the `naming polynomial' i.e. 1 for $s$, $n_x$ for $p_{n_x}$, $n_xn_y$ for $d_{n_xn_y}$ 
etc. 
(Note that the name `$z^2$' in $d_{{n_z}^2}$ is indeed {\sl shorthand} for $z^2$ -- 1/3; shorthand begins 
to proliferate if one goes beyond the d-orbitals; the polynomials arising in our working are also 
subject to being `nonunique' under $\sum_{i = 1}^3\{n^i\}^2 = 1$.)  
That the wavefunctions are their own naming polynomials is via Sec 2.4's analogy 2 mirroring how the 
orbitals in space historically got their Cartesian names, and also is akin to representations \cite{CH} 
of the spherical harmonics in terms of homogeneous polynomials.  
Another form for the solution 
is\footnote{This is  
found by shifting from arctan to arccos and then using one of the standard definitions of Tchebychev 
polynomials, T$_{\td}(\xi) = \mbox{cos(\d}\,\mbox{arccos}(\xi))$   
Despite being the product of two generally nonpolynomial factors, the two conspire to produce 
polynomials in each case.  
We then introduce the symbol ${\cal T}_{\td}(\xi)$ to mean $T_{\td}(\xi)$ for cosine solutions and 
                             $\sqrt{1 - T_{\td}(\xi)^2}$                  for sine solutions.}
\beq
\Psi_{\sD|\sd|}(n^i) \propto \mP^{\sd}_{\sD}(n_z){\cal T}_{\sd}
\left(
n_x/\sqrt{{n_x}^2 + {n_y}^2} 
\right)  = 
\mP^{\sd}_{\sD}(\mbox{RelSize(12,34)}){\cal T}_{\sd}
\left(
\mbox{RelSize(1,2)}/\sqrt{1 - \mbox{RelSize(12,34)}^2} 
\right)
\mbox{ } .  
\eeq
However, via Sec 2.3's tessellation trick, we can interpret the wavefunctions in terms of the metroland 
mechanics on the sphere itself, on which they take the particularly familiar `orbital' form.

For D, d = 0, 0 ($s$-orbital), note that the axis is arbitrary so it is evident from using 2 different 
principal axes that the probability distribution function on-axis is not to be trusted in spherical 
coordinates about that axis. 
We conclude that the ground state does not have bias toward any particular configurations.  
\noindent For  D, d = 1, 0 ($p_{n_z}$ orbital), equatorial configurations are improbable, meaning that 
mergers of the \{12\} and \{34\} clusters [including the non \{12,34\} DD's] are disfavoured, while 
polar configurations are probable, meaning that the \{12\} and \{34\} clusters being small and 
well-apart is favoured.
\noindent For D, d = 1, $\pm$1, in the $p_{n_y}$ orbital case, the $n_y$ = RelSize(3,4) axis part of the 
equator is probable, so mergers of a small \{12\} and a large \{34\} are favoured.  
$p_{n_x}$ is the $\{12\} \leftrightarrow \{34\}$ of this. 
\noindent For D, d = 2, 0 ($d_{{n_z}^2}$ orbital), both equatorial and polar configurations are 
probable, so that the \{12\} and \{34\} clusters are either merged or small and well-apart.  
\noindent For D, d = 2, $\pm$1, in the $p_{n_yn_z}$ case, equatorial configurations are improbable, so 
mergers of \{12\} and \{34\} are improbable, and also the \{12\} cluster is small;  DD's are disfavoured.  
$d_{n_xn_z}$ is the $\{12\} \leftrightarrow \{34\}$ of this. 
\noindent For D, d = 2, $\pm$2, in the $d_{{n_x}^2 - {n_y}^2}$ case, equatorial configurations, i.e. 
mergers of \{12\} and \{34\}, are probable, especially those with one but not both of the clusters 
are large (i.e. configurations along one of the RelSize (12) or RelSize(34) axes: contents 
inhomogeneity), including the \{13,24\}, \{14,23\}, \{23,14\} and \{24,13\} DD's.
In the ${d_{n_xn_y}}$ case, again equatorial configurations are probable, but now with $|\mbox{RelSize}(12)| 
\approx |\mbox{RelSize}(3,4)|$ i.e. contents homogeneity, including the DD's where the two clusters are  
on top of each other [\{13,24\} and \{14,23\}].

\subsection{Overlap integrals: shapes and spreads of shape operators}

We are interested furthermore in computing overlap integrals 
$\langle \mD_1 \d_1| \widehat{\mbox{Operator}} | \mD_2 \d_2\rangle$ for three applications 
1) expectation and spread of shape operators (below).  
2) Time-independent perturbation theory about the very special solution in Sec 3.5.  
3) Time-dependent perturbation theory on space of shapes with respect to a time provided by the scale in 
the shape-scale split ERPM models in semiclassical formulation also makes use of these. 
This parallels Halliwell--Hawking's work \cite{HallHaw} and embodies one of our program's 
eventual goals, so we prefer giving details of computing the overlaps to giving details of 2).  
2) and 3) have the merit of extending to far more general potential terms than the harmonic oscillator-like terms 
discussed in the present working, while 2) survives as a subproblem in the corresponding 
time-independent non-semiclassically approximated shape-scale split ERPM.

The idea to use 1) such can be traced back to how expectations and spreads of powers of $r$ are used in 
the study 

\noindent of atoms (see e.g. \cite{Messiah, Schwinger} for elementary use in the study of hydrogen, or 
\cite{FF} for use in approximate studies of larger atoms).   
Doing this amounts to acknowledging that `modal' quantities (peaks and valleys), as read off from plots 
or by the calculus, are only part of the picture: such as the mean  
$\langle\mn\,\ml\,\mm\,|\,r\,|\,\mn\,\ml\,\mm \rangle$, 
$\langle\mn\,\ml\,\mm\,|\,r^2\,|\,\mn\,\ml\,\mm\rangle$  
and the spread $\Delta_{\sn\,\sll\,\sm}(r) = \sqrt{\langle\mn\,\ml\,\mm\,|\,r^2\,|\,\mn\,\ml\,\mm\rangle 
- \langle\mn\,\ml\,\mm\,|\,r\,|\,\mn\,\ml\,\mm\rangle^2 }$ are also useful.  
E.g. for hydrogen, one obtains from the angular factors of the integrals trivially cancelling and 
orthogonality and recurrence relation properties of Laguerre polynomials in Appendix B for the radial 
factors that
\beq
\langle\mn\,\ml\,\mm\,|\,r\,|\,\mn\,\ml\,\mm\rangle = \{3\mn^2 - \ml\{\ml + 1\}\}a/2 \mbox{ } \mbox{ and } \mbox{ } 
\Delta_{\sn\,\sll\,\sm}r = 
\sqrt{\{\mn^2\{\mn^2 + 2\} - \{\ml\{\ml + 1\}\}^2\}}a/2 \mbox{ } ,
\eeq
where $a$ is the Bohr radius.
One can then infer from this that a minimal typical size is $3a/2$ and that 
the radius and its spread both become large for large quantum numbers.  
c.f. how the modal estimate of minimal typical size is $a$ itself; the slight 
disagreement between these is some indication of the limited accuracy to which either estimate should be 
trusted. 
Also, we identify the above as expectations of scale operators, and thereby next ask whether they 
have pure shape counterparts in the standard atomic context.

The answer is yes.  Up to normalization, they are the 3-$Y$ integrals \cite{LLQM} (for $Y$ spherical 
harmonics, the radial parts of the integration now trivially cancelling), and the general case of 
this has been evaluated in terms of Wigner 3j symbols \cite{LLQM}. 
Furthermore, many of the integrals for the present paper's specific cases of interest are written out case-by 
case in \cite{Mizushima} (this applies to expectations of the RelSize's as well as $B$, $C$, $D$, 
$E$, $F$ perturbation terms' constituent overlaps).  
Shape operators for hydrogen are also considered in \cite{AF} (briefly) and \cite{LSB}.

Moreover, the context in which shape operators occur in molecular physics is wider than just the above.  

\noindent
E.g. 1) expectations of cos$\beta$ for $\beta$ a relative angle from inner products between physically 
meaningful vectors e.g. between the 2 electron--nucleus relative position vectors in Helium, in the 
characterization of molecules' bonds or in nuclear spin-spin coupling (p 443 of \cite{AF}).  

\noindent 
E.g. 2) one also gets expectations of $Y_{20}(\theta)$ [c.f. form 4 of (\ref{39})] in spin-spin and 
hyperfine interactions (p 437-441) of \cite{AF} (as a shape factor occurring alongside a $1/r^3$ scale factor.  

\noindent
E.g. 3) In the study of the $H_2^+$ molecular ion, one uses fixed nuclear separation as a scale setter 
and then one has not only 1 relative angle but also 2 ratios forming spheroidal coordinates with respect 
to which this problem separates, and expectations of all these things then make good sense.  

\noindent
We contemplate `mini- and midi'superspace counterparts of such shape operators in Sec 4.5.  


As regards good shape operators for 4-stop metroland, the kinematical quantization 
carries guarantees that the three $n^i$ are promoted to good quantum operators.  
These can be interpreted as RelSize(1,2), RelSize(3,4) and RelSize(12,34) as per Sec 2.3.      
It is also useful to note at this stage that $n_z$ is not only physically RelSize(12,34) but also 
mathematically the Legendre variable.

Then $\langle\mD\,\md\,|\,\widehat{\mbox{RelSize(1,2)}}\,|\,\mD\,\md\rangle = 
       \langle\mD\,\md\,|\,\widehat{\mbox{RelSize(3,4)}}\,|\,\mD\,\md\rangle = 0$ and 
$\langle\mD\,\md\,|\,\widehat{\Theta}\,|\,\mD\md \rangle \approx$
$\langle\mD\,\md\,|\,\widehat{\mbox{RelSize(12,34)}}\,|\,\mD\,\md\rangle = 0$   
as an obvious result of orientational symmetry. 
%
The useful information starts with the spreads,  
\beq
\Delta_{\sD\,\sd}(\widehat{\mbox{RelSize(1,2)}})  = 
\sqrt{\frac{\mD\{\mD + 1\} + \md^2 - 1}{\{2\mD - 1\}\{2\mD + 3\}}Q_1(\d)}
\mbox{ } \mbox{ } , \mbox{ } \mbox{ }
\Delta_{\sD\,\sd}(\widehat{\mbox{RelSize(3,4)}})  = 
\sqrt{\frac{\mD\{\mD + 1\} + \md^2 - 1}{\{2\mD - 1\}\{2\mD + 3\}}Q_2(\d)} \mbox{ } ,
\eeq
\beq
\Delta_{\sD\,\sd}(\widehat{\Theta}) \approx \Delta_{\sD\sd}(\widehat{\mbox{RelSize(12,34)}})  = 
\sqrt{\frac{2\{\mD\{\mD + 1\} - \md^2\} - 1}{\{2\mD - 1\}\{2\mD + 3\}}} \mbox{ } ,
\eeq
for $Q_2(\d) =         1/2 \mbox{ for the } \d \mbox{ cosine solution, }  
                       3/2 \mbox{ for the } \d \mbox{ sine solution, and } 
                       1   \mbox{ otherwise,}$ 
and $Q_1(\d)$ the sin $\leftrightarrow$ cos of this. 
One can then readily check that 
$\langle \widehat{n}_x\mbox{}^2 + \widehat{n}_y\mbox{}^2 + \widehat{n}_z\mbox{}^2\rangle$ = 1, 
as it should be.

One case of interest is the ground state.
Therein, the spreads in each are $1/\sqrt{3}$.  
Another case of interest is the large quantum number limit.  
$\Delta_{\sD\,\sd}(\widehat{\Theta}) \approx$ 
$\Delta_{\sD\,\sd}(\widehat{\mbox{RelSize(12,34)}})$ which, for the maximal d ($|\md| = \mD$), 
is equal to $1/\sqrt{2\mD + 3}$ which goes as $1/\sqrt{2\mD} \longrightarrow 0$ for D large.  
The hydrogen counterpart of this result is $\Delta_{\sll\,\sll}\hat{\theta} \approx  1/\sqrt{2\ml} 
\longrightarrow 0$, i.e. restriction to the Kepler--Coulomb plane (e.g. \cite{AF} outlines this, while 
\cite{LSB} considers it in more detail).   
Back to our problem, this result therefore signifies recovery of the equatorial classical geodesic as 
the limit of an ever-thinner belt in the limit of large maximal projectional relative dilational 
quantum number $|\md| = \mD$ (`the rim of the disc' of Sec 2.10, traversed in either direction 
according to the sign of d).   
In fact, as for the constant potential we can put the axis wherever we please, this leads to recovery 
of {\sl any} of the classical geodesics.  
Also, for d = 0, $\Delta_{\sD\,0}\widehat{\Theta} \approx 
\Delta_{\sD\,0}(\widehat{\mbox{RelSize(12,34)}}) \longrightarrow 1/\sqrt{2}$ for D large.  
This means that the $s$, $p_{n_z}$, $d_{{n_z}^2}$ ... sequence of orbitals does not get much narrower 
as D increases, so that for these states we only get limited peaking about clusters \{12\} and \{34\} 
both being small and well apart, a situation which we will revisit in the next subsection due to its 
centrality to the assumptions made in, and applications of, this paper.
The RelSize(1,2) and RelSize(3,4) operators' spreads tend to finite constant values for large D no 
matter what value d takes.

What of $\hat{\Phi}$?  
Now, clearly, by factorization and cancellation of the $\Theta$-integrals, the d = 0 states obey the 
uniform distribution over 0 to $2\pi$, with mean $\pi$ and variance $\pi^2/3$ (corresponding to 
axisymmetry).  
Furthermore, $\langle\mD\,\md\,|\,\hat{\Phi}\,|\,\mD\,\md\rangle$ is also $\pi$ and cosine and sine 
states have 
\beq
\Delta_{\sD\,\sd}(\widehat{\Phi}) = \sqrt{{\pi^2}/{3} + {1}/{2\md^2}} 
\mbox{ } \mbox{ and } \mbox{ } 
\Delta_{\sD\,\sd}(\widehat{\Phi}) = \sqrt{{\pi^2}/{3} - {1}/{2\md^2}} \mbox{ } ,
\eeq 
which indicate some resemblance to the uniform distribution arising for large d (mean and 
variance do not see the multimodality, but at least, by inspection along the lines of the 
preceding subsection, it is {\sl regular} multimodality for d maximal -- equatorial flowers of 2D 
petals -- by inspection of the shapes of the standard maximal $s$, $p$, $d$, $f$, $g$ ... orbitals. 


\subsection{Solution in special case -- large and small regimes}

Passing to stereographic coordinates, banal-conformal transforming to the flat representation and 
applying the small approximation, our Schr\"{o}dinger equation becomes 
\beq
-\{\hbar^2/2\}
\big\{
{\cal R}^{-1}
\{
{\cal R}\Psi_{,{\cal R}}
\}_{,{\cal R}} 
+ {\cal R}^{-2}\Psi_{,\Phi\Phi}
\big\} = {\cal E} - {\omega^2{\cal R}^2}/{2} \mbox{ } ,
\eeq
which is in direct correspondence with the 2-$d$ quantum isotropic harmonic oscillator (see e.g. 
\cite{Messiah, Schwinger, Robinett} under ${\cal R} \longleftrightarrow r$ (radial coordinate), $1 
\longleftrightarrow$ particle mass, and with our $\omega$ as classical frequency ($\times I$).   
Thereby,  
\beq
{\cal E} = \mbox{\Large n}\hbar\omega \mbox{ } \mbox{ for } \mbox{ } 
\mbox{\Large n} \equiv 1 +  2\sN + |\d| 
\label{zin}
\eeq
for $\sN$ a node-counting quantum number running over $\mathbb{N}_0$ and d a `projected' dilational 
quantum number as in the preceding subsection but now running over $\mathbb{Z}$.
[The `shifted energy' in its usual units, $\fE^{\prime} = \fE - A - B$, itself goes as 
\beq
\fE^{\prime} = \{\mbox{\Large n}^2\hbar^2/2\}\{1 + \sqrt{1 - B\{4/\mbox{\Large n}\hbar\}^2}\}  
\mbox{ } ,
\eeq
so for $\mbox{\Large n}\hbar/\omega << 1$ (small quantum numbers as used below), $\fE^{\prime}I 
\approx \mbox{\Large n}\hbar\Omega$ for $\Omega = 2\sqrt{-B}$.] 
%
%
The solutions are then (to suitable approximation) 
\beq
\Psi_{\tN\sd}(\Theta, \Phi) \propto {\Theta}^{|\sd|}\{1 + {|\d|\Theta^2}/{12}\}\mbox{exp}
(-{\omega\Theta^2}/{8\hbar})\mL_{\tN}^{|\sd|}({\omega\Theta^2}/{4\hbar})\mbox{exp}(\pm i\md\Phi)
\label{gi}
\eeq
for $\mL_{a}^b(\xi)$ the associated Laguerre polynomials in $\xi$ (see Appendix B).  
[The $\Phi$-factor of this is rewriteable as before in terms of the $n^i$ or RelSize(12,34) and 
RelSize(1,2), while the $\Theta$-factor is now a somewhat more complicated function of RelSize(12,34)].

The large regime gives the same eigenvalue condition (\ref{zin}), and (\ref{gi}) again for wavefunctions 
except that one now uses the supplementary angle $\Xi = \pi - \Theta$ in place of $\Theta$.   
Next, see Fig 5 for the form and interpretation of the wavefunctions.  


{             \begin{figure}[ht]
\centering
\includegraphics[width=0.5\textwidth]{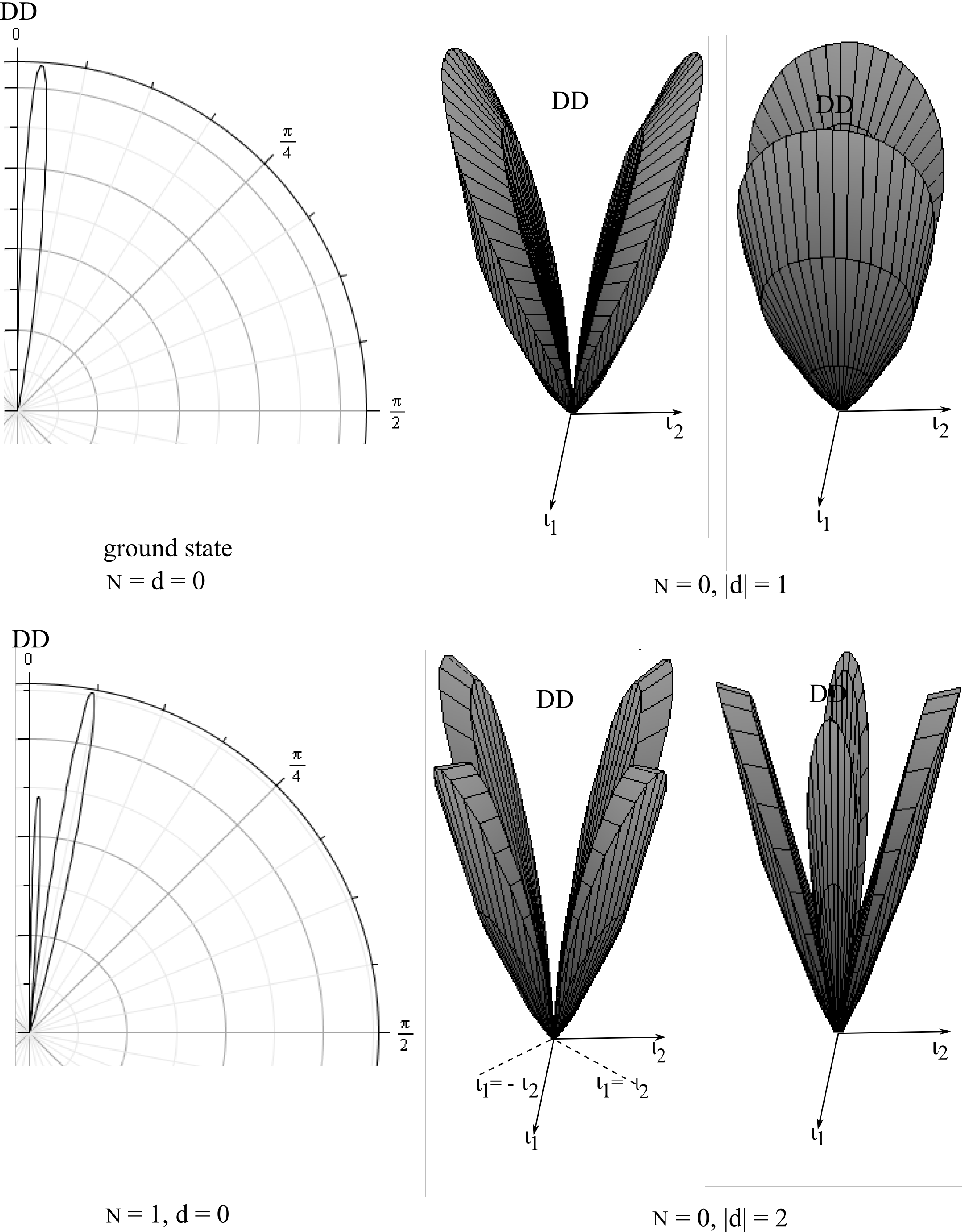}
\caption[Text der im Bilderverzeichnis auftaucht]{        \footnotesize{    Probability density 
functions for this subsection's problem for $\omega/\hbar$ large, 400, say, plotted using Maple 
\cite{Maple}.  
All d = 0 states are axisymmetric about the \{12,34\} clustering's DD axis, i.e. all relative sizes 
for cluster \{12\} and for cluster \{34\} are equally favoured.  
The ground state is peaked around the \{12,34\} DD collision.  
It it the surface of revolution of the given curve.  
The $\sN$ = 0, $|\d|$ = 1 solutions are a degenerate pair.  
Each takes the form of a pair of inclined lobes -- the cosine solution's oriented about the $n_x =$ 
RelSize(1,2) = 0 D collision and the sine solution's about the $n_y =$ RelSize(3,4) = 0 D collision.
These next three solutions form a degenerate triplet.
The $\sN = 1$, d = 0 solution is a slender bulge around the \{12,34\} DD collision, then a gap and then 
a second bulge in the form of a cone, representing `a band very close to this DD collision and a band 
somewhat close to it being probable, while all other configurations are improbable.  
The $\sN$ = 0, $|\d|$ = 2 solutions are tulips of four petals, the cosine one separately favouring the 
lunes of near \{12\} D collisions and near \{34\} D collisions (i.e. contents inhomogeneity),  while 
the sine one disfavours these and favours instead the lunes at $\pi/4$ to the preceding, which 
correspond to contents homogeneity of the \{12\} and \{34\} clusters.  

The large case's approximate solution is just the reflection of the preceding about the equatorial 
plane with the same interpretation except that \{34\} is now to the left of \{12\}.    }        }
\label{Fig5}
\end{figure}            }


In the small regime, the RelSize(1,2) and RelSize(3,4) operators still have zero expectation as each 
sign for these remains equally probable.  
For D, d substantially smaller than $\omega/\hbar$ so powers of the latter dominate powers of the 
former (and $\omega/\hbar$ {\sl was} considered to be large, so this works for the kind of quantum 
numbers in this subsection's specific calculations), the following mean and spread results for shape 
operators are derived using orthogonality of, and a recurrence relation for, Laguerre polynomials, as 
provided in Appendix B.  
\beq
\langle\sN\,\md\,|\,\widehat{\mbox{RelSize(12,34)}}\,|\,\sN\,\md\rangle = 
1 - 2\mbox{\Large n}\hbar/{\omega} \mbox{ } . 
\eeq
$\Delta_{\tN\,\sd}\widehat{\mbox{RelSize(12,34)}}$ is zero to first two orders, beyond which the 
approximations used begin to break down, but it would appear to have leading term proportional to 
$\hbar/\omega$.  
These results signify that the potential has trapped what was much more uniform in Sec 3.3 into 
a narrow area around the \{12,34\} DD collision.  
Furthermore, 
\beq
\Delta_{\tN\,\sd} \widehat{n^{\bar{a}}} \approx 
\sqrt{2\mbox{\Large n}\hbar Q_{\bar{a}}(\d)/\omega}   
\eeq 
for $\bar{a} = 1, 2$ gives $\Delta_{\tN\,\sd} (\widehat{\mbox{RelSize(1,2)}})$ for $\bar{a} = 1$ and 
$\Delta_{\tN\,\sd} (\widehat{\mbox{RelSize(3,4)}})$ for $\bar{a} = 2$.
So one can obtain strong concentration around the poles by suitable choice of springs, amounting to a tall thick 
equatorial barrier and polar wells.  
The ground state has the tightest spread in RelSize(1,2) and RelSize(3,4): $\sqrt{2\hbar/\omega}$.  
This has some parallels with how the Bohr radius is an indicator of atomic size, including the 
hydrogen--isotropic harmonic oscillator correspondence \cite{Schwinger}.


\subsection{Perturbations about the very special solution}

We begin by recasting our Schr\"{o}dinger equation in Legendre variables $n_z = \mbox{cos}\,\Theta$, 
\beq
\{
\{1 - n_z\mbox{}^2\}\Psi_{,n_z} 
\}_{,n_z} 
+ \{1 - n_z\mbox{}^2\}^{-1}\Psi_{,\Phi\Phi} = \{{\cal A} - {\cal B} + n_z\mbox{}^2\{2{\cal B} - 
{\cal C}\mbox{cos}\,2\Phi\} + {\cal C}\mbox{cos}\,2\Phi\}\Psi \mbox{ } .  
\label{yt}
\eeq
One can then study this using time-independent perturbation theory (see e.g. \cite{LLQM} for derivation 
of the formulae for this up to second order).  
Applying perturbation theory here means considering 
1) $C$ small, which is high contents homogeneity at the level of each cluster's (Hooke 
cooefficient)/(reduced mass) in the sense that $K_1 - K_2$ is small compared to $\hbar^2$.
2) $B$ small, in the sense that $\hbar^2$ is large compared to $\{K_1 + K_2\}/2 - K_3$, which 
collapses to $K_1 - K_3$ small in the case of $C = 0$, meaning that there is little difference between 
the inter-cluster spring and the intra-cluster springs.

Perturbative study of (\ref{yt}) is amenable to exact calculations though involving various of 
trigonometric and standard/tabulated associated Legendre function integrals, or, alternatively, 
the aforementioned 3-Y integrals. 
Furthermore, this continues to be the case if one includes a non-diagonal/non-normal basis' 
$D$, $E$ and $F$ terms.

For the ${\cal B}$-perturbation, as both it and the unperturbed Hamiltonian commute with ${\cal D}$, the 
eigenvalue problem can be solved separately in each subspace ${\cal V}_{\sd}$ of a given eigenvalue $\d$ 
of ${\cal D}$, and in each such subspace the spectrum of the unperturbed Hamiltonian is nondegenerate, so 
that nondegenerate perturbation theory is applicable  (this argument parallels e.g. p 697 of 
\cite{Messiah}).   
This gives (with the unperturbed problem's ${\cal A}$ playing the role usually ascribed to the energy 
and $\fH_1$ the perturbative term)    
${\cal A}_{\sD\,\sd}^{(1)} = \langle\mD\,\md\,|\,{\fH}_1\,|\,\mD\,\md\rangle$  
at first order and 
$
{\cal A}_{\sD\,\sd}^{(2)}$ =--$\sum_{\sD^{\prime},\sd^{\prime} \neq \sD,\sd} 
      |\langle\mD^{\prime}\md^{\prime}|\,{\fH}_1\,|\,\mD\,\md\rangle|^2/
      \{ {\cal A}_{\tD^{\prime}} - {\cal A}_{\tD}    \} \mbox{ }   
$
at second order \cite{LLQM}.
Then e.g. \cite{08II} double use of a standard recurrence relation \cite{AS} gives a $\Delta \md = 0$, 
$\Delta \mD = 0, \pm 2$ `selection rule'.
Moreover, the terms that survive this take the following forms.    
\beq
\langle\mD\,\d\,|\,{\cal B}\{2n_z\mbox{}^2 - 1\}\,|\,\mD\,\d\rangle = 
{{\cal B}\{1 - 4\d^2\}}/{\{2\mD - 1\}\{2\mD + 3\}}
\mbox{ } ,  
\label{*****}
\eeq
which is closely related to the expectation of $n_z$ = RelSize(12,34) already computed in Sec 3.3. 
Two new overlaps that 

\noindent are more general than expectations are 
\beq
\langle\mD + 2\,\d\,|\,{\cal B}\{2n_z\mbox{}^2 - 1\}\,|\,\mD \,\d\rangle = \frac{2{\cal B}}{2\mD + 3}
\sqrt{\frac{\{\{\mD + 2\}^2 - \d^2\}\{\{\mD + 1\}^2 - \d^2\}}{\{2\mD + 5\}\{2\mD + 1\}}}
\mbox{ } , 
\label{***}
\eeq
and then, swapping D for D -- 2, also,  
\beq
\langle\mD - 2\,\d\,|\,{\cal B}\{2n_z\mbox{}^2 - 1\}\,|\,\mD\,\d\rangle = \frac{2{\cal B}}{2\mD - 1}
\sqrt{\frac{\{\{\mD^2 - \d^2\}\{\{\mD - 1\}^2 - \d^2\}}{\{2\mD + 1\}\{2\mD - 3\}}}
\mbox{ } . 
\label{****}
\eeq
Using these then gives the perturbed `energies':  
$$
\fE_{\sD\,\sd} = A + {\hbar^2}\mD\{\mD + 1\}/2 + {B\{1 - 4\md^2\}}/{\{2\mD - 1\}\{2\mD + 3\}} + 
$$
\beq
\frac{4B^2\{ \{2\mD + 5\}\{2\mD + 3\}^3\{\mD^2 - \md^2\}\{\{D - 1\}^2 - \md^2\}  - 
             \{2\mD - 1\}^3\{2\mD - 3\}\{\{\mD + 2\}^2 - \md^2\}\{\{\mD + 1\}^2 - \md^2\}\}\}}
     {\hbar^2\{2\mD + 5\}\{2\mD + 3\}^3\{2\mD + 1\}\{2\mD - 1\}^3\{2\mD - 3\}} + O(B^3).
\label{big}
\eeq
Note that d positive and negative are treated the same, so there is only a partial uplifting of degeneracy.  
Changes to the wavefunction due to the perturbations for the sign of $B$ corresponding to Sec 3.4 and to 
second order in $B$ are that we get slight bulges at the poles for the ground state (a bit of 
$d_{{n_z}^2}$ mixed in).  


The ${\cal C}$ perturbation can likewise be studied based on half-way stage overlaps that can be directly 
transcribed by our angular momentum to dilational momentum analogy from those computed in e.g. 
\cite{Mizushima}.  
E.g. the surviving terms are found \cite{08II} by a second standard recurrence relation  
\cite{AS} to obey the selection rule $\Delta\md = \pm 2$, $\Delta \mD = 0$.   
Some noteworthy features of the study of the ${\cal C}$ term are that degenerate perturbation theory is 
now required, there is no first order contribution as $\Delta \d = \pm 2$ only, and now d and --d do get 
shifted differently corresponding to this perturbation not preserving the axis of symmetry.   
In nondiagonal/nonnormal form, the further $D$ term has the same selection rule to the $C$ term's 
while the $E$ and $F$ terms share the selection rule $\Delta \d = \pm 1$, $\Delta \mD = 0, \pm 2$.  
The above `noteworthy features' apply to these also.


\subsection{Molecular physics analogies} 

Analogy A) (\ref{chi}) occurs in mathematical physics (e.g. from the separation of the wave equation in 
prolate spherical coordinates \cite{Brief, SMCH, MorseFeshbach, AS}) and has multiple applications in 
molecular physics studies of which parallel some of the studies in the present paper.    
Examples of this in molecular physics are as follows.  


\noindent Analogy A.1) (\ref{chi}) recast in terms of the Legendre variable is 
\beq
\{
\{1 - {n_z}^2\}\Psi_{,n_z} 
\}_{,n_z} 
- \{1 - {n_z}^2\}^{-1}\mm^2\Psi = \{{\cal A} - {\cal B} + 2{\cal B}{n_z}^2\}\Psi 
\mbox{ } ,  \label{ytt}
\eeq
which is the easier of the two spheroidal equations that arise in the study of the $H_2^+$ molecular ion 
\cite{BLS, Slater, H2+}.    
This and the next two analogies are for $B < 0$, although the aforementioned mathematical physics 
literature covers $B > 0$.


\noindent
Analogy A.2) The potential $V_0\{1 - \mbox{cos}\,2\theta\}$ [c.f. form 3 of (\ref{39})] occurs in 
modelling the rotation of a linear molecule in a crystal \cite{P,SPW,WS}.  
Here, the analogy is (\ref{ana2},\ref{ana3},\ref{ana4}) where the axis and rotor in question are 
provided by the linear molecule itself, `energy' $\leftrightarrow$ energy up to a constant, 
\beq
K_1/2 \leftrightarrow 2V_0 \mbox{ up to the same constant difference as in the energy analogy }
\eeq
\beq
B \leftrightarrow -2V_0
\eeq

\noindent Analogy A.3) The potential $-\alpha_{||}\ttE^2\,\mbox{cos}^2\theta$ [c.f.form 2 of (\ref{39})] 
for $\alpha_{||}$ the polarizability along the axis occurs in the study \cite{C,TS} of e.g. the $CO_2$ 
molecule in a background electric field $\ttE$ (the study of polarizability is the theory underlying 
Raman spectroscopy).     
Here the analogy is, rather,
\beq
B \longleftrightarrow - {\alpha_{||}}\ttE^2/2 \mbox{ } . 
\eeq

\noindent 
Analogy A.4) Examples 2) of Sec 3.3 is another substantially developed area in the molecular physics 
literature. 


\noindent 
Analogy B) is with the ammonia molecule $NH_3$, in the following rougher but qualitatively valuable 
sense.   
$NH_3$ has two potential wells separated by a barrier and then is capable of tunnelling between the two 
at the quantum level (like an umbrella inverting in the wind).  
Our model for $B < 0$ is similar to this, albeit in spherical polar coordinates: we have 2 polar wells 
with an equatorial barrier in between.  

\noindent
This analogy then gives us some idea about how the separate solutions for the two wells compose. 
For $NH_3$, one can start with separate solutions for each well and additional degeneracies ensue (due 
to the wells being identical and being able to distribute some fixed energies between these in diverse 
ways). 
However the wavefunctions tend to perturb each other toward breaking these degeneracies, forming 
symmetric and antisymmetric wavefunctions over the two wells \cite{AF, Slater}.

\subsection{Applications of the analogies and developing an overall picture of our model}

Firstly, the $B < 0$ locally stable small or large regimes are the kind of regimes that are termed 
`rotator-like' in analogy A.2)'s literature; both of the SO(3) quantum numbers (for us, dilational 
quantum numbers) hold good in this regime.

Secondly, for $B < 0$ one can use Analogy B to form a simple picture of putting the small and large 
$\Theta$ approximations together.  
As d remains a good quantum number for the unapproximated problem, one expects to need the North Pole  
approximation's d and the South Pole approximation's d to match and the subsequent perturbations exacted 
by these two approximations upon each other not to affect d.  
Also, prior to any recombination, one has degeneracies as follows (call the near-North Pole's 
node-counting quantum number $\sN$ and the near-South Pole's $\sN^{\prime}$).
There is the one ground state    $\sN = \sN^{\prime} = \d = 0$, 
then the degenerate pair       $\sN = \sN^{\prime} = 0, \d = \pm 1$, and 
then the degenerate quadruplet $\sN = 1, \sN^{\prime} = 0$ 
                            or $\sN = 0, \sN^{\prime} = 1$ for each of  $\d = \pm 1$.   
Now if $\sN$ and $\sN^{\prime}$ match, expectation of RelSize(12,34) goes to 0 again though the 
wavefunction's distribution is bimodal about both poles. 
If they do not match, RelSize(12,34) retains some nonzero expectation due to the peaking near the two 
poles being different in detail.  
The flip here, as in $NH_3$, is an inversion, i.e. it reverses the orientation, sending 1,2,3,4 to 
4,3,2,1.

Thirdly, analogy A.3) is well-known for its Raman-type $\pm 2$ and not $\pm 1$ selection 
rule, which parallels our results of Sec 3.5.
Analogy A.3) has furthermore been studied perturbatively for what for us is the small $B < 0$ regime.  
This allows us to e.g. check the half-way house results (\ref{*****}, \ref{***}, \ref{****}) against 
p 271-273 of \cite{TS}.

Fourthly, further resources from analogy A.1)'s references \cite{SMCH, MorseFeshbach} include analysis 
of this equation's poles in the complex plane and how it admits a solution in the form of an infinite 
series in associated Legendre functions in the vicinity of $\pm 1$ and in Bessel functions in the 
vicinity of $\infty$, as well as how to piece together these different representations. 
%
%
It is then appropriate to compare results from the expansion in associated Legendre functions against 
our perturbative regime (this particular working holds regardless of the sign of $B$).
Thus we find the lowest four cases of (\ref{big}) to agree with p 1502-4 of \cite{MorseFeshbach}, which 
additionally provides the corresponding wavefunctions which we use to first order in $B$ in Sec 4.4 to 
evaluate the na\"{i}ve Schr\"{o}dinger interpretation probabilities for these states' model universes 
being large.

Fifthly, one cannot really put together our near-polar calculations and our perturbative calculations,  
because the ``$B$ small perturbative condition" goes a long way toward $\omega$ being small and then 
only a bit of the wavefunction is near the pole.   
Our near-polar calculations should be compared, rather, with the asymptotics for $B$ large.    
Analogy A.2)'s literature covers this for what for us is the $B$ large negative ($>> \fE - A - B = 
\fE^{\prime}$) regime, giving, via the analogy, 
\beq
\fE^{\prime} \mbox{ } \widetilde{\mbox{ }} \mbox{ } \mbox{\Large n}\hbar\omega_{\sll\sa\sr\sg\se}
+ O(1/\omega_{\sll\sa\sr\sg\se})
\mbox{ } \mbox{ } \mbox{and}   
\label{HOlike}
\eeq 
\beq
\Psi \propto \mbox{exp}(\omega_{\sll\sa\sr\sg\se}\mbox{cos}\Theta/\hbar)
\{\{\mbox{tan$\frac{\Theta}{2}$}\}^{2\tN}\{\mbox{sec$\frac{\Theta}{2}$}\}^{2\{|\sd| + 1\}} 
+ O(1/\omega_{\sll\sa\sr\sg\se}) \} \mbox{ } 
\eeq
as the relevant asymptotic solutions, for $\omega_{\sll\sa\sr\sg\se} = 2\sqrt{-B}$.  
Now, from (\ref{oc},\ref{qus}) the small-$\Theta$ approximate solution (\ref{gi})'s 
$\omega = 8\sqrt{-B} = 4\omega_{\sll\sa\sr\sg\se}$, so $\mbox{exp}(\omega_{\sll\sa\sr\sg\se}\mbox{cos}\Theta/\hbar)$ in (\ref{HOlike}) $\approx 
\mbox{const}\times\mbox{exp}(\{\omega/4\hbar\}\{-\Theta^2/2\})$, which is indeed in agreement with 
the leading and dominant factor of (\ref{gi}).
For our model, this regime signifies that $K_3 >> K_1, K_2$ i.e. that the inter-cluster spring is 
much stronger than each of the intra-cluster springs.    
This is termed a `harmonic oscillator-like regime' -- comparing (\ref{HOlike}) and the standard result 
for the 2-$d$ isotropic harmonic oscillator makes it clear why.   
d alone is a good dilational quantum number in this regime.

Sixthly, the spheroidal equation has led to many hundreds of pages of tabulations \cite{SMCH} and further 
numerical work e.g. in \cite{AS, +numerics}, though the most recent of this states that this study 
is still open in some aspects.

Seventhly, one can furthermore envisage extending analogy A.2) to have the further parallel with our 
model that a rotationally-dislocated molecule in a cubic crystal will have preferred directions in space 
of approximately the same form as ours are in configuration space.  
We do not know if such a study has been done.

Finally, we comment that A.3) has been extended \cite{+Raman} to include what for us are $C$, $D$, $E$ 
and $F$ terms. 
For, what one has more generally is a symmetric polarization tensor $\alpha$ such that $\mu_{\rho} = 
\alpha_{\rho\sigma}\ttE_{\sigma}$.  
Then for the $CO_2$ model in a diagonal basis $\alpha_z = \alpha_{||}$ giving the combination 
-$\alpha_{\perp}\,\mbox{sin}^2\theta - \alpha_{||}\,\mbox{cos}^2\theta$ [a slight improvement of analogy 
A.3) by inclusion of the smaller $\alpha_{\perp} = \alpha_x = \alpha_y$], and this readily rearranges to 
the special case of the third form of (\ref{39}).  
But for more general groups than just oxygen atoms at each end of the axis (while still remaining in a 
diagonal basis) $\alpha_x \neq \alpha_y$, giving $-\alpha_{x}\,\mbox{sin}^2\theta\,\mbox{cos}^2\phi - 
\alpha_{y}\,\mbox{sin}^2\theta\,\mbox{cos}^2\phi - \alpha_{z}\,\mbox{cos}^2\theta$ which is the general 
case of the second form of (\ref{39}).  
Moreover, in non-diagonal bases, the off-diagonal elements form extra terms directly analogous to those 
in (\ref{3More}).  
Thus there is an extended analogy between our problem and the study of polarization, with 4-stop 
metroland's Jacobi--Hooke coefficients forming a configuration space-indexed analogue of the 
spatial-indexed polarizability tensor.

\section{Conclusion}

Relational particle models (RPM's) benefit from notions of locality and structure that are absent in 
minisuperspace and are free of many of the technical difficulties of midisuperspace models. 
This makes them suitable for testing some of the conceptual aspects of quantum cosmology, and of quantum 
general relativity (such as the Problem of Time).  
In particular, in this paper we study the RPM of 4 particles in 1-$d$ -- {\it 4-stop metroland} -- in the 
case without scale, both classically and quantum-mechanically.  
We concentrate on the clustering into two particular binary clusters [of particles \{12\} and particles 
\{34\}] both by using coordinates that follow this case and imposing a potential term that restricts the 
physics to being near such a configuration.  
This is toward a qualitative conceptual model of the quantum cosmological seeding of structure formation 
in a semiclassical regime (paralleling the Halliwell--Hawking \cite{HallHaw} approach, which is somewhat 
narrower as a Problem of Time strategy but has further conceptual and computational applications outside 
of the Problem of Time context too), and of records theory \cite{PW83, GMH, B94II, EOT, H99, Records}.  
The counterpart of the current paper's model with scale (which is harder and in which the current 
paper's work occurs as a subworking under the shape--scale split) will be required for some aspects of 
such a study (in particular, for a semiclassical treatment with a greater number of parallels 
to that of GR).  
This is further work in progress \cite{MGM, Cones, scaleQM, SemiclIII}, though Sec 4.2--4 give a brief 
account of generalizations of the current paper's model and how these meet additional quantum 
cosmological and Problem of Time criteria.

This paper's model has an $\mathbb{S}^2$ configuration space and then the mathematics which follows has 
analogies with the standard axisymmetric sphere and central force problems of ordinary mechanics.
In particular, where a conserved angular momentum occurs in these analogue problems, a conserved 
relative dilational momentum occurs in our model. 
[These both have SO(3) mathematics, but each has a different physical nature, the two being embraced by 
our notion of rational momentum which generalizes angular momentum to ratios that do not happen to 
physically be angles.]  
We then interpret some of 4-stop metroland's classical and quantum solutions in cases with harmonic 
oscillator-like potentials.  
The solutions in spherical variables give fairly standard mathematics such as that of the rigid rotor 
and of the 2-$d$ isotropic harmonic oscillator in some of the simpler cases, albeit now these require 
subsequent interesting and unusual interpretation in terms of the 4-stop metroland problem's mechanical 
variables. 
We deduce this at the level of mass-weighted coordinates by tessellating the shape space sphere by the 
mechanical interpretation appropriate to 4-stop metroland, which we find to possess the symmetry group 
of the cube.  
Further tools we introduce, paralleling basic treatises on the atom, are expectations and spreads of 
shape operators, to which we can also attribute cosmological analogies.    
Our shape operators are RelSize(12,34): the relative size difference between universe and its 
\{12\}, \{34\} cluster contents, 
RelSize(1,2): the size of the \{12\} cluster relative to the size of the whole model universe, 
and its \{34\} cluster counterpart.  
The polar angle $\Phi$ itself is an inhomogeneity ratio of the contents of the universe themselves 
(i.e. of the two clusters relative to each other).

We obtain expectations and spreads for such operators e.g. in ground state and in large quantum number 
limits.  
We consider the very special constant potential case that resides within the harmonic oscillator-like 
potential models as a particularly structurally homogeneous balance of springs, as well as more general 
cases treated perturbatively for small differences in spring constitution, asymptotically for large 
such differences, and in near-polar approximations.  
We further benefit from recognizing that the special case with the two clusters of the same spring 
constitution but the inter-cluster spring is weaker gives an elsewise well-known spheroidal 
equation, alongside various molecular physics analogies: with $H_2^+$, $NH_3$, 
rotation of molecules in crystals and molecular polarizability (at least the last of which extends to 
cases with more general combinations of springs between the four particles).  
This permits us to tap into substantial mathematical physics results and generally control our 
particular problem, and is a further example of useful bridges between RPM quantum cosmology 
models and the physics of molecules (triangleland RPM with harmonic oscillator-like potentials having 
already been found to share mathematics with the Stark effect for a linear rigid rotor \cite{08II}).

\subsection{Comments on extension to N $>$ 4 metrolands}

The Jacobi H- and K-coordinates of Fig 1 generalize to cover an increasing variety of `part H-shaped, 
part K-shaped' clusterings (\cite{ACG86} may be useful in this respect), which are of additional value 
as less trival models of structure formation and of records theory.
For full reduction for scalefree arbitrary-N-stop metroland and the subsequent Euler--Lagrange equations 
in the arbitrary-potential case, see \cite{FORD, 08I}.  
Moreover, we now comment that the number and nature of conserved quantities that each of these possesses 
is tied to the usual SO(N -- 1) representation theory, and their physical interpretations extend the 
present paper's discovery of dilational quantities.   
Tessellations by physical interpretation are now harder as they both have more pieces and also are 
more difficult to visualize due to being higher-dimensional.  
However, our relative size and contents inhomogeneity shape operators do straightforwardly extend to 
N-stop metroland.

Within each N-stop metroland, one can envisage a tower of special, very special, ... (very)$^{\sN - 2}$ 
special problems.  
The most special of these in each case has a constant potential and thus gives ultraspherical geodesics 
classically and the ultraspherical rigid rotor quantum-mechanically (solved by ultraspherical harmonics 
\cite{08II}), while the next most special of these in each case has (N -- 1)-$d$ isotropic harmonic oscillator 
mathematics in its near-polar regime (solved by a power times a Gaussian times an associated 
Laguerre polynomial).
Establishing a perturbative regime about each most special problem would then appear to be possible e.g. 
\cite{08II} by recurrence relations of the Gegenbauer polynomials \cite{AS,GrRy}.  
One technical difference is that, if one does use conformal operator ordering, then one can no longer 
use the configuration space being 2-$d$ to evoke collapse to Laplacian ordering like in Sec 3.1. 
However, hyperspheres are of constant curvature and so of constant Ricci scalar curvature, so 
$\xi\mbox{Ric}(\mathbb{S}^k)$ is just a constant, $\xi k\{k - 1\}$ (our `$\fE$ has no nonconstant 
prefactors banal conformal representation' having the unit sphere as its configuration space). 
So, even in this case, the sole difference between Laplace and conformal ordering (or any other member 
of the $D^2 - \xi$Ric($M$) family of operators) is in what is to be interpreted to be the zero of the 
energy.  
We also note that for N = 5 the analogy with the Halliwell--Hawking scheme is somewhat tighter, as both 
involve perturbative expansions in $\mathbb{S}^3$ ultraspherical harmonics.  
Finally, the next most special equation unapproximated can also be mapped to the spheroidal equation, 
so that the fairly standard mathematical physics of that equation continues to be of aid in N-stop 
metroland.

\subsection{Comments on Extension to metrolands with scale} 

The configuration spaces for these are cones over the corresponding shape spaces \cite{Cones}.  
The present paper's advances in the physical understanding of conserved quantities in RPM's have 
further applications here.
Our introduction of shape quantities to be promoted to operators also continues to be relevant here  
through there being a shape--scale split, so that evaluating shape operators here collapses back to 
pure shape workings such as the present paper's.    
Expectation and spread of size (in close parallel with atomic physics) will also now be pertinent.   
Metrolands with scale will now have solid rather than surface analogues of the present paper's 
`tessellation by physical interpretation' technique.
The way in which similarity RPM arises as a subproblem from the shape-scale split of the scaled theory 
means that the most special harmonic oscillator case and perturbations thereabout survives as a piece of 
the analysis upon introduction of scale, now partnered by isotropic harmonic oscillators in the size 
quantity.

This setting with scale is more appropriate as regards both toy-modelling of cosmology in general and 
in particular to using a semiclassical approach both in the Problem of Time context and in the 
Halliwell--Hawking context. 
One of us makes a first sketch at this in the smallest case in \cite{MGM}, with other cases to follow in 
\cite{Cones, scaleQM, SemiclIII}.

\subsection{Comments on extension to 2-$d$}  

The price to pay in introducing scale is that one no longer has nontrivial constraints in spatially 
1-$d$ models with which to model some effects due to GR's momentum constraint.      
On the long term, one can get around this by passing to spatial dimension $> 1$.  
A first such model is triangleland: the RPM of 3 particles in the plane.   
This has a $\mathbb{S}^2$ shape space like the present paper's 4-stop metroland model does. 
This gives a number of useful insights. 
E.g. parts of the present paper parallel \cite{08I, 08II}. 
Even more significantly, because the 4-stop metroland interpretation of the sphere turns out to be 
more straightforward, the present paper allows for an improved understanding of the less straightforward 
triangleland case \cite{+Tri}.
Scaled triangleland is harder; so far we have just provided some classical study for this (also to 
be augmented by the present paper's techniques at the classical level in \cite{Cones} toward finally 
providing a quantum study of it in \cite{08III}).

Moreover, triangleland lacks the present paper's nice feature of splitting into two nontrivial 
subsystems (of 2 particles each), which is a useful nontriviality from the structure formation and 
records theory perspectives.  
Studying scaled quadrilateralland (RPM of 4 particles in the plane) \cite{QShape} would incorporate this 
feature too.
Thus this model would possess a number of midisuperspace's features with the benefit of being 
technically simpler.  
This makes it particularly suitable for the simultaneous investigation of records theory and the 
semiclassical approach (which may support each other, and histories theory, to form a more robust 
combined approach to the Problem of Time and to quantum cosmology \cite{Halliwell, H99, 08II}). 
Quadrilateralland does have a further technical complexity -- its shape space is $\mathbb{CP}^2$, which 
unavoidably involves complex-projective mathematics (triangleland has $\mathbb{CP}^1$ but this is 
well-known to also be $\mathbb{S}^2$).

Further features for consideration in RPM models involve \cite{Ultra} A) oriented shapes -- real 
projective spaces $\mathbb{RP}^{\sN - 2}$ in place of $\mathbb{S}^{\sN - 2}$ as shape spaces or  
$\mathbb{CP}^{\sN - 2}/\mathbb{Z}_2$ in place of 
$\mathbb{CP}^{\sN - 2}$ as shape spaces and the corresponding cones in models with scale and/or 
B) (partial) particle indistinguishibility by which only pieces of whichever of the preceding spaces 
would pass to being the configuration spaces.  
The present paper's treatment of physical interpretation by multiple coordinate charts and 
tessellations by physical interpretation are doubtlessly ideas of further value in the study of these 
models with their wide range of configuration space geometries.

\subsection{Further details of Problem of Time applications} 

Our wavefunctions, eigenvalues and operators are useful in the following Problem of Time investigations. 

\noindent
1) The computation of na\"{\i}ve Schr\"{o}dinger interpretation  \cite{HP86UW89} probabilities of the 
universe having some particular property.  

\noindent
Example 1) consider quantifying P(universe is large), in the sense that the two clusters under study 
are but specks in the firmament, by P($\epsilon$-close to the \{12,34\} double-double collision), which 
means, at the level of the configurations themselves, that the magnitude of $\sqrt{\mbox{RelSize}(1,2)^2 
+ \mbox{RelSize}(3,4)^2}/$Relsize(12,34) lies between 1 and 1 -- $\epsilon^2/2$, and, in configuration 
space terms, that one is in the $\epsilon$-caps about each pole.  
Then from the latter and by the na\"{\i}ve Schr\"{o}dinger interpretation, this probability $\propto$
$\int_{\epsilon\mbox{\scriptsize --caps}}|\Psi|^2\d S = 
 \int_{\Phi = 0}^{2\pi}\{\int_{\Theta = 0}^{\epsilon} + \int_{\Theta = \pi - \epsilon}^{\pi}\}
|\Psi(\Theta, \Phi)|^2\mbox{sin}\,\Theta\,\d\Theta\d\Phi$.  
So, e.g. for the very special solution's ground state and first excited state, one gets proportionality 
to $\epsilon^2 + O(\epsilon^4)$, while for the states with dilational quantum 

\noindent numbers D = 1, $|$d$|$ = 1, 
one gets proportionality to $\epsilon^4$ + $O(\epsilon^6)$.  

\noindent
Example 2) consider quantifying P(the two clusters nominally under study are in fact merged) 
by P($\delta$-close to \{12,34\} merger) which means, at the level of the configurations themselves, that 
the size of Relsize(12,34) does not exceed the small number $\delta$, and, in configuration space terms, 
that one is in the $\delta$-belt around the equator.  
Thus 

\noindent 
P($\delta$-close to \{12,34\} merger) $\propto \int_{\delta\mbox{\scriptsize --belt}}|\Psi|^2\d S$, 
which, in the very special case, works out to be proportional to $\delta^3 + O(\delta^5)$ for D = 1 
d = 0 and to $\delta + O(\delta)^3$ for the other three lowest-lying states.

\noindent 
Example 3) consider quantifying P(universe is contents-homogeneous) in the sense that the 
two clusters under study are similar to each other, by the magnitude of 
RelSize(1,2)/RelSize(3,4) departing from 1 by no more than $2\eta$.  
Then, on configuration space, one is in the tetralune described in Fig 2, and the 
na\"{\i}ve Schr\"{o}dinger interpretation gives 

\noindent P(universe is $\eta$-contents-homogeneous)  
$\propto \int_{\eta\mbox{\scriptsize --tetralune}}|\Psi|^2\d S$, which, in the very special case, 
comes out as proportional to $\eta$ for all four of the lowest-lying states.

Example 1 also makes sense for the small-regime special solution. 
One now obtains proportionality to $\epsilon^2 \sqrt{\omega/\hbar}$ to leading order i.e. the same 
`(small)$^2$' factor as in the very special problem but now with an opposing `$\sqrt{\mbox{large}}$' 
factor, amounting to the small regime's potential well (Fig 3b) concentrating the wavefunction near 
the poles i.e. in the region of the configuration space corresponding to large universes in the 
above-described sense.

Finally, also repeating Example 1 for the wavefunctions with first order perturbative corrections in $B$ 
included, we now find proportionality to 
$\epsilon^2\{1 - 8BI^2/9\hbar^2  \} + O(B^2) + O(\epsilon^4)$ for the ground state, to 
$\epsilon^2\{1 - 8BI^2/25\hbar^2 \} + O(B^2) + O(\epsilon^4)$ for D = 1, d = 0, and 
$\epsilon^4\{1 - 36I^2B/25\hbar^2\} + O(B^2) + O(\epsilon^6)$ for the D = 1, d = 1 states.  
The signs of these corrections conform with intuition, as (Fig 3a) $B > 0$ corresponds to placing 
a potential barrier at the poles and a well around the equator, which should indeed decrease the amount 
of wavefunction there, i.e. making large universes less probable, and vice versa for $B < 0$.    


\noindent
2) Given explicit wavefunctions such as this paper's, one can build up projectors and mixed states 
(including with environment portions traced out) and then construct conditional probabilities \cite{PW83} for pairs 
of universe properties. 

\noindent
3) As regards records theory \cite{Records}, the current paper's classical work provides some means of 
defining a notion of distance on configuration space (which is quite closely related to the measure 
problem in cosmology \cite{GHS87HP88GT08}), and a notion of localizability in space.  
Next \cite{RPMRec}, one would construct notions of information (alias negentropy) both at the classical 
level and at the 

\noindent
quantum level for the problems solved in this paper.  
In this respect it is worth noting that QM perturbation theory suffices in order to build an approximate 
statistical mechanics \cite{LLSM}.
Such notions of information include e.g. Shannon's, von Neumann's, Tsallis's \cite{Records}, as well as 
notions of subsystem information, mutual information and correlation (such as the covariance for the 
two clusters in the situation that the present paper centres on).     

\noindent 
4) One can also build up decoherence functionals for histories theory \cite{Hartle}, and consider the  
Feynman--Vernon influence functional that Halliwell uses \cite{H99} for the study of records within 
histories theory.  

\noindent 
5) Exact wavefunctions also serve as useful checks on whether the semiclassical approach's assumptions 
and approximations are appropriate \cite{scaleQM, SemiclIII}.  

\noindent N.B. Operator insertions for meaningful shape operators remain useful in constructing various of 
these Problem of Time-relevant objects.  
All of 2) to 5) above workings being substantially longer than 1), we leave them for future occasions.

\subsection{Analogues of our shape operators in mini and midisuperspace?}

Another longer-term goal would be to export insights acquired by our program to 
`mini and midi'superspace; are there then useful analogues of shape operators for these (anisotropy 
operators, inhomogeneity operators?)  
In surveying the literature, we have found, firstly, that Kucha\v{r} and Ryan \cite{KR89} consider 
$\langle y^2 \rangle$ for $y$ a reparametrization of one of the anisotropy degrees of freedom 
$\beta_{\pm}$ in diagonal Bianchi IX quantum cosmology.
%
%
Secondly, Ashtekar and Bojowald make mention of an anisotropy operator in studying loop quantum gravity 
\cite{AB}.  
Thirdly, Petryk and Schleich \cite{PS} consider expectation values for geometrical quantities in the 
Hartle--Hawking initial state in their study of conditional probabilities in the 3-$d$ Ponzano-Regge 
minisuperspace.  
Fourthly, Halliwell and Hawking \cite{HallHaw} compute the expectation of the anisotropy in temperature 
of the microwave background; this has the additional value of being ``halfway to midisuperspace" in that 
it considers inhomogeneous perturbations about a homogeneous spacetime. 
As regards inhomogeneous spacetimes, the Lema\^{\i}tre--Tolman--Bondi solution principally concerns 
radial scale variables, so it is far more of an analogue to the shape-scale extension of the 
present paper.  
While there is not anything as yet that we know about using shape operators at the quantum 
level for the Gowdy universe, e.g. Andersson, van Elst and Uggla \cite{AVU} use a form of shape--scale 
variables at the classical level.  
Thus, while shape quantities and shape operators have occasionally been used in `mini and 
midi'superspace, a systematic treatment parallelling that in atomic physics does as yet appear to be 
lacking.    

\mbox{ }

\noindent{\bf Acknowledgments}:
%
\noindent
we thank Claire Anderson for hospitality, and Professors Don Page, Jonathan Halliwell and Julian Barbour 
for references and discussion. 
This research was partly supported by Grant Number RFP2-08-05 from The Foundational Questions Institute 
(fqxi.org). 

\mbox{ }

\noindent{\bf\large Appendix A This paper's generalization of angular momentum put into context}

\mbox{ } 

\noindent
What is habitually called angular momentum mathematics (because that is a common guise in which it appears in physics), 
is, de facto, the representation theory of SO(p).   
The most usual case is that of SO(3), habitually called the rotation group, and also interpretable as 
the isometry group of the 2-sphere [more generally, SO(p) is the p-dimensional rotation group and the  
isometry group of the (p -- 1)-sphere].  
In turn, SO(3) is closely related to SU(2) (which is its double cover).  
The rational momentum viewpoint is more general than the mechanical angular momentum perspective but not the SO(p) 
one, as Fig 6 begins to explain by classifying examples.  
Beyond that, one can get SO(p) as a piece of an even larger group. 
[This is already the case for Fig 6's Runge--Lenz example, which can be viewed as a second 
SO(3) coming from a partly-`hidden' SO(4), but it covers further cases where the group is not 
necessarily `hidden', such as SO(3, 1) in relativistic particle physics, or SU(3) containing 3 
different directions' worth of SU(2) ladder operators].  
Quantum `SO(p) objects' also combine under an addition rule, whereby composites of cases in Fig 6 arise.  
E.g. total angular momentum T = L + S in atomic physics; more generally, total rational momentum 
${\cal T} = {\cal R} + {\cal A}$, which also would include the effect of adding internal `arrows' to 
the present paper's 1-$d$ RPM setting.\footnote{As  
the word `spin' itself has rotational and hence angular momentum connotations, we generalize `spin 
alias internal angular momentum', S, to `arrow alias internal rational momentum', ${\cal A}$.}
%
There are also rational momenta that are linear combinations of relative angular momentum and relative 
dilational momenta occur in triangleland \cite{+Tri}.
Finally, there are SO(p) quantities that are physically phase space objects.  
In this sense, this paper's rational momentum came about from building a different notion of 1-$d$ space 
from the usual into the present paper (scalefreeness)  
Finally, adding arrows to the present model could be nontrivial in the sense that the 1-$d$ arrows need 
not just obey separate `tensored-on' occupation rules since dilational momentum exists as does addition 
of arrow and dilational momentum quantities, so that spatially 1-$d$ models can have arrow-dilational 
momentum interactions that parallel the spin-orbital angular momentum couplings that occur in higher 
spatial dimensions. 

{            \begin{figure}[ht]
\centering
\includegraphics[width=0.5\textwidth]{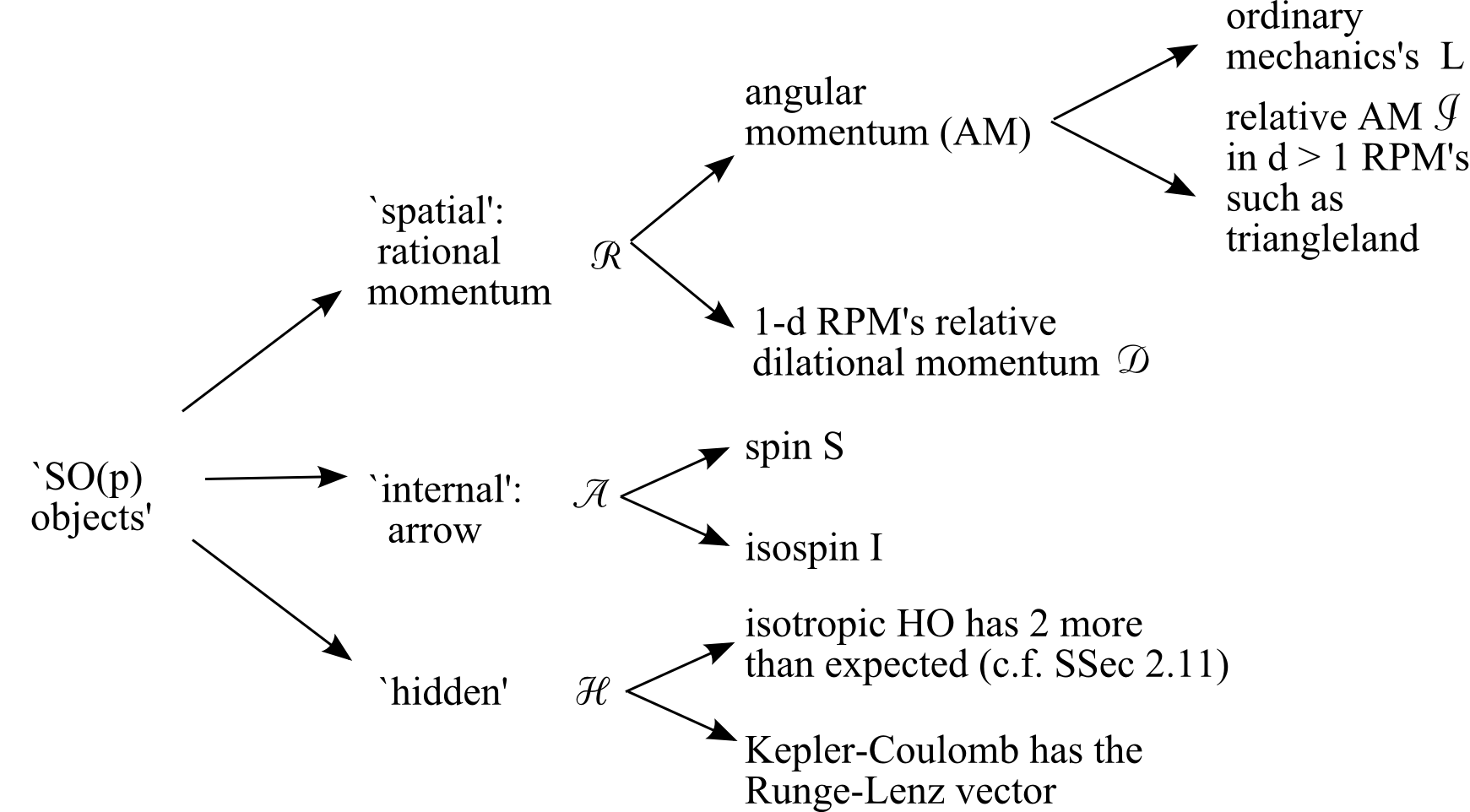}
\caption[Text der im Bilderverzeichnis auftaucht]{        \footnotesize{Various physical 
realizations of SO(p) objects.    }        }
\label{Fig6}
\end{figure}            }


\noindent{\bf\large Appendix B Some results concerning associated Laguerre Polynomials}

\mbox{ } 

\noindent  
The bounded solutions of the associated Laguerre equation 
\beq
\xi y_{,\xi\xi} + \{\alpha + 1 - \xi\}y_{,\xi} + \beta y = 0
\eeq 
are the associated Laguerre polynomials $\mL_{\beta}^{\alpha}(\xi)$. 
These obey \cite{AS} the orthogonality relation

\noindent
\beq
\int_0^{\infty}\xi^{\alpha}\mbox{exp}(- \xi)L^{\alpha}_{\beta}(\xi)L^{\alpha}_{\beta^{\prime}}(\xi)
\d\xi = 0 \mbox{ unless } \beta = \beta^{\prime} 
\eeq
and the recurrence relation
\beq
\xi\mL^{\alpha}_{\beta}(\xi) = \{2\beta + \alpha + 1\}\mL_{\beta}^{\alpha}(\xi) - 
                               \{\beta + 1\}          \mL_{\beta + 1}^{\alpha}(\xi) - 
                               \{\beta + \alpha\}     \mL^{\alpha}_{\beta - 1}(\xi) \mbox{ } .  
\eeq


\end{document}